\documentclass[twocolumn]{aastex62}
\usepackage{amsmath}
\usepackage{graphicx}
\usepackage{mathtools} 
\DeclarePairedDelimiter{\abs}{\lvert}{\rvert}

\newcommand{\HII}{\mbox{H II}}
\newcommand{\NII}{\mbox{[N II]}}
\newcommand{\SII}{\mbox{[S II]}}
\newcommand{\Ha}{H$\alpha$}
\newcommand{\NIIHa}{\NII/H$\alpha$}

\newcommand{\ktd}{KMOS$^{\rm{3D}}$}
\newcommand{\sfrsd}{$\Sigma_{\rm SFR}$}
\newcommand{\vout}{$v_{\rm out}$}
\newcommand{\rout}{$R_{\rm out}$}
\newcommand{\smsd}{$\Sigma_*$}
\newcommand{\kms}{\mbox{km s$^{-1}$}}
\newcommand{\myrkpc}{M$_\odot$~yr$^{-1}$~kpc$^{-2}$}

\shorttitle{A Kiloparsec Scale View of Star-Formation Driven Outflows at $z\sim$~2.3}
\shortauthors{Rebecca L. Davies et al.}

\begin{document}

\title{Kiloparsec Scale Properties of Star-Formation Driven Outflows at $z\sim$~2.3 in the SINS/zC-SINF AO Survey}\thanks{Based on observations collected at the European Organisation \\ for Astronomical Research in the Southern Hemisphere under \\ ESO Programme IDs 075.A-0466, 079.A-0341,  080.A-0330, \\ 080.A-0339, 080.A-0635, 081.A-0672, 081.B-0568, 183.A-0781,\\ 087.A-0081, and 088.A-0209.}

\correspondingauthor{Rebecca L. Davies} 
\email{rdavies@mpe.mpg.de}

\author{R.L. Davies}
\affiliation{Max-Planck-Institut f\"ur extraterrestrische Physik,
             Giessenbachstrasse, D-85748 Garching, Germany}

\author{N.M. F\"orster Schreiber}
\affiliation{Max-Planck-Institut f\"ur extraterrestrische Physik,
             Giessenbachstrasse, D-85748 Garching, Germany}
             
\author{H. \"Ubler}
\affiliation{Max-Planck-Institut f\"ur extraterrestrische Physik,
             Giessenbachstrasse, D-85748 Garching, Germany}
             
\author{R. Genzel}
\affiliation{Max-Planck-Institut f\"ur extraterrestrische Physik,
             Giessenbachstrasse, D-85748 Garching, Germany}
\affiliation{Department of Physics, Le Conte Hall,
             University of California, Berkeley, CA 94720}
\affiliation{Department of Astronomy, Hearst Field Annex,
             University of California, Berkeley, CA 94720}

\author{D. Lutz}
\affiliation{Max-Planck-Institut f\"ur extraterrestrische Physik,
             Giessenbachstrasse, D-85748 Garching, Germany}

\author{A. Renzini}
\affiliation{INAF-Osservatorio Astronomico di Padova,
             Vicolo dell'Osservatorio 5, I-35122 Padova, Italy}

\author{S. Tacchella}
\affiliation{Harvard-Smithsonian Center for Astrophysics,
             60 Garden Street, Cambridge, MA 02138, USA}

\author{L.J. Tacconi}
\affiliation{Max-Planck-Institut f\"ur extraterrestrische Physik,
             Giessenbachstrasse, D-85748 Garching, Germany}

\author{S. Belli}
\affiliation{Max-Planck-Institut f\"ur extraterrestrische Physik,
             Giessenbachstrasse, D-85748 Garching, Germany}
             
\author{A. Burkert}
\affiliation{Max-Planck-Institut f\"ur extraterrestrische Physik,
             Giessenbachstrasse, D-85748 Garching, Germany}
\affiliation{Universit\"ats-Sternwarte, Ludwig-Maximilians-Universit\"at, 
	    Scheinerstr. 1, M\"unchen, D-81679, Germany}
             
\author{C.M. Carollo}
\affiliation{Institute of Astronomy, Department of Physiscs,
             Eidgen\"ossische Technische Hochschule,
             Z\"urich, CH-8093, Switzerland}
             
\author{R.I. Davies}
\affiliation{Max-Planck-Institut f\"ur extraterrestrische Physik,
             Giessenbachstrasse, D-85748 Garching, Germany}

\author{R. Herrera-Camus}
\affiliation{Max-Planck-Institut f\"ur extraterrestrische Physik,
             Giessenbachstrasse, D-85748 Garching, Germany}

\author{S.J. Lilly}
\affiliation{Institute of Astronomy, Department of Physiscs,
             Eidgen\"ossische Technische Hochschule,
             Z\"urich, CH-8093, Switzerland}

\author{C. Mancini}
\affiliation{Dipartimento di Fisica e Astronomia, Universit\`a di
          Padova, Vicolo dell'Osservatorio 2, I-35122 Padova, Italy}
\affiliation{INAF-Osservatorio Astronomico di Padova,
             Vicolo dell'Osservatorio 5, I-35122 Padova, Italy}
             
\author{T. Naab}
\affiliation{Max-Planck-Institut f\"ur Astrophysik, 
	    Karl-Schwarzschildstr. 1, D-85748 Garching, Germany}
             
\author{E.J. Nelson}
\affiliation{Max-Planck-Institut f\"ur extraterrestrische Physik,
             Giessenbachstrasse, D-85748 Garching, Germany}
\affiliation{Harvard-Smithsonian Center for Astrophysics,
             60 Garden Street, Cambridge, MA 02138, USA}

\author{S.H. Price}
\affiliation{Max-Planck-Institut f\"ur extraterrestrische Physik,
             Giessenbachstrasse, D-85748 Garching, Germany}

\author{T.T. Shimizu}
\affiliation{Max-Planck-Institut f\"ur extraterrestrische Physik,
             Giessenbachstrasse, D-85748 Garching, Germany}
             
\author{A. Sternberg}
\affiliation{School of Physics and Astronomy, Tel Aviv University, 
	    Tel Aviv 69978, Israel}
             
\author{E. Wisnioski}
\affiliation{Research School of Astronomy \& Astrophysics, Australian National University, Canberra, ACT 2611, Australia}
\affiliation{ARC Centre for Excellence in All-Sky Astrophysics in 3D (ASTRO 3D), Australia}

\author{S. Wuyts}
\affiliation{Department of Physics, University of Bath,
             Claverton Down, Bath, BA2 7AY, UK}

\begin{abstract}
We investigate the relationship between star formation activity and outflow properties on kiloparsec scales in a sample of 28 star forming galaxies at $z\sim$~2~--~2.6, using adaptive optics assisted integral field observations from SINFONI on the VLT. The narrow and broad components of the \Ha\ emission are used to simultaneously determine the local star formation rate surface density (\sfrsd), and the outflow velocity \vout\ and mass outflow rate $\dot{M}_{\rm out}$, respectively. We find clear evidence for faster outflows with larger mass loading factors at higher \sfrsd. The outflow velocities scale as \mbox{\vout~$\propto$ \sfrsd$^{0.34 \pm 0.10}$}, which suggests that the outflows may be driven by a combination of mechanical energy released by supernova explosions and stellar winds, as well as radiation pressure acting on dust grains. The majority of the outflowing material does not have sufficient velocity to escape from the galaxy halos, but will likely be re-accreted and contribute to the chemical enrichment of the galaxies. In the highest \sfrsd\ regions the outflow component contains an average of $\sim$45\% of the \Ha\ flux, while in the lower \sfrsd\ regions only $\sim$10\% of the \Ha\ flux is associated with outflows. The mass loading factor, $\eta$~=~$\dot{M}_{\rm out}$/SFR, is positively correlated with \sfrsd\ but is relatively low even at the highest \sfrsd: $\eta \lesssim$~0.5~$\times$~(380~cm$^{-3}$/n$_e$). This may be in tension with the \mbox{$\eta$ $\gtrsim$~1} required by cosmological simulations, unless a significant fraction of the outflowing mass is in other gas phases and has sufficient velocity to escape the galaxy halos. 
\end{abstract} 


\section{Introduction}
\nocite{NMFS18a}
\nocite{Newman12_406690}

Galaxy scale outflows are expected to play a major role in regulating the star formation and chemical enrichment histories of galaxies \citep[e.g.][]{Dave12, Hopkins12, Vogelsberger13, Hirschmann13, Chisholm17}, mediating the co-evolution of galaxies and their central supermassive black holes \citep[e.g.][]{Silk98, Fabian12, King15}, and setting the sizes of galaxy disks \citep[e.g.][]{Okamoto05, Sales10}. Powerful winds driven by star formation and AGN activity transfer large amounts of mass and energy from galaxies to the surrounding circumgalactic medium \citep[e.g.][]{Peeples14, Tumlinson17}, depleting the supply of cold gas available for star formation within galaxies and preventing the circumgalactic gas from cooling and falling back onto galaxies \citep[e.g.][]{DiMatteo05, Springel05, Croton06, Hopkins06, Somerville08, Erb15, Beckmann17}. Outflows are therefore thought to play an important role in driving the low baryon fractions of galactic disks \citep{Dekel86, Efstathiou2000, Sales10}. The cosmic stellar mass density peaks at $\sim$20\% of the cosmic baryon density for a stellar mass of $\log(M_*/M_\odot$)~$\sim$~10.5, and drops to \mbox{$\sim$5-10\%} towards higher stellar masses (where black hole accretion feedback is most efficient) and lower stellar masses (where star formation feedback is most efficient) \citep[e.g.][]{Baldry08, Moster13, Moustakas13, Behroozi13}.

Star formation driven outflows are expected to have the biggest impact on galaxies at $z\sim$~1~--~3, during the peak epoch of star formation \citep[see][and references therein]{Madau14}. Blueshifted absorption components are ubiquitous in the rest-frame UV spectra of $z~\sim$~2 star forming galaxies, revealing extended (tens of kiloparsec) reservoirs of diffuse outflowing material expelled from galaxies over long periods of time \citep[e.g.][]{Shapley03, Weiner09, Rubin10, Steidel10, Erb12, Kornei12, Bordoloi14}. Broad high velocity components in the rest-frame optical emission line spectra of star forming galaxies trace denser outflowing material within a few kiloparsecs of the launching points of the outflows. These emission components provide an instantaneous snapshot of the current outflow activity and are seen in $\sim$~10-30\% of star forming galaxies at $z~\sim$~2 \citep[when AGN host galaxies are explicitly excluded;][]{Genzel11, Newman12_global, Freeman17, NMFS18b}. Despite the prevalence of star formation driven outflows at high redshift, there are few quantitative constraints on their physical properties.

Many studies have reported trends between the velocities of star formation driven outflows and the global $M_*$, SFR and/or \sfrsd\ of their host galaxies \citep[e.g.][]{Rupke05, Martin05, Weiner09, Rubin10, Steidel10, Erb12, Kornei12, Newman12_global, Talia12, Arribas14, Bordoloi14, Rubin14, Chisholm16, Heckman16, Sugahara17, NMFS18b}. Correlations between the outflow velocity \vout\ and star formation properties arise naturally because the level of star formation activity determines the amount of energy injected into the ISM by supernovae, stellar winds and radiation pressure from massive stars \citep[e.g.][]{Chevalier85, Strickland04, Murray11}. The stellar feedback combines with the turbulence driven by disk instabilities to counteract the disk gravity and launch outflows \citep[see e.g.][]{Ostriker11, Krumholz18}. The galaxy stellar mass drives the depth of the local potential which decelerates the outflowing material, but is also positively correlated with the SFR \citep[e.g.][]{Brinchmann04, Elbaz07, Noeske07, Peng10, Whitaker14} which determines the amount of energy available to accelerate the outflowing material.

It is important to accurately characterise the SFR-\vout\ and \sfrsd-\vout\ relationships, because their scalings provide constraints on the primary outflow driving mechanism(s). If the outflows are driven by mechanical energy from supernovae and stellar winds, the outflow velocity is predicted to be weakly dependent on the level of star formation activity (\vout~$\propto$~\sfrsd$^{0.1}$; \citealt{Strickland04, Chen10}, \vout~$\propto$~SFR$^{0.2-0.25}$; \citealt{Ferrara06, Heckman00}). On the other hand, if the outflows are radiatively driven, the outflow velocity is predicted to scale strongly with the level of star formation (\vout~$\propto$~\sfrsd$^2$; \citealt{Murray11, Kornei12}, \vout~$\propto$~SFR; \citealt{Sharma12}). If the dominant outflow driving mechanism varies within individual galaxies, the power law scaling will be intermediate between the energy and momentum driven cases.

The slopes of the \sfrsd-\vout\ and SFR-\vout\ relations remain a matter of debate. Some studies report relatively flat power law scalings with indices of 0.1-0.15 \citep[e.g.][]{Chen10, Arribas14, Chisholm16}, while other studies report somewhat steeper scalings with power law indices of 0.25-0.35 \citep[e.g.][]{Martin05, Weiner09, Heckman16, Sugahara17}. The discrepancies between different scalings reported in the literature are likely to originate from differences in the observed outflow tracers, adopted definitions of \vout, and range of probed outflow velocities (see discussions in e.g. \citealt{Kornei12, Heckman17}). 

The relationship between the level of star formation activity and the incidence and properties of outflows has, for the most part, only been investigated using galaxy integrated values. However, high spatial resolution observations of high SFR galaxies (both at $z~\sim$~2 and in the local universe) indicate that star formation driven outflows are launched from small ($\sim$~1~kpc) regions coincident with the most intense star formation events \citep{Shopbell98, Genzel11, Newman12_406690, Bolatto13}. Therefore, it may not be the global level of star formation which is most relevant for shaping the outflow properties, but the \textit{local} level of star formation. \citet{Bordoloi16} found that the properties of the outflowing material along different lines of sight to a lensed galaxy at $z~\sim$~1.7 are correlated with the properties of the nearest star forming region, suggesting that the outflows are indeed `locally sourced'. On the other hand, \citet{James18} found that the strongest outflow in a lensed galaxy at $z~\sim$~2.38 is associated with the most diffuse star forming region, suggesting that the outflow is `globally sourced'. Systematic studies of larger galaxy samples are required to conclusively determine whether the properties of star formation driven outflows are more strongly dependent on global or local galaxy properties.

In this paper, we investigate the relationship between resolved $\sim$1-2~kpc scale (0.15-0.25'') star formation activity and the incidence and properties of outflows in a sample of 28 star forming galaxies at $z\sim$~2.3 from the SINS/zC-SINF AO Survey \citep{NMFS18a}. We build on the work of \citet{Newman12_global}, who explored the relationship between \textit{global} galaxy properties and the incidence and velocity of outflows in a similar sample of galaxies to the one used in this paper, and the work of \citet{Genzel11} and \citet{Newman12_406690}, who performed detailed analyses of the star formation and outflow properties of individual star forming clumps in 5 SINS/zC-SINF AO galaxies, of which 3 are included in our sample. Here we extend these analyses to study outflow properties as a function of resolved physical properties across 28 galaxies, considering not only the highly star forming clump regions but also the less active inter-clump regions.

The outline of the paper is as follows. We probe the star formation and outflow properties on $\sim$0.17'' scales using adaptive optics assisted integral field observations of the \Ha\ emission line (described in Section \ref{sec:sample}). We stack the spectra of individual spaxels of the integral field datacubes to create high signal-to-noise (S/N) spectra in bins of resolved physical properties, and perform single and multi-component emission line fitting to analyse the properties of the outflow component in individual stacks (described in Section \ref{sec:sins_method}). We explore which physical properties are most closely linked to the presence of outflows in Section \ref{sec:broad_emission}, and present more detailed results on the relationship between the local \sfrsd\ and the incidence and properties of outflows in Section \ref{subsec:outflow_properties}. The implications of our findings are discussed in Section \ref{sec:discussion} and our conclusions are presented in Section \ref{sec:conc}.

Throughout this work we assume a flat $\Lambda$CDM cosmology with \mbox{H$_{0}$ = 70 \kms\ Mpc$^{-1}$} and \mbox{$\Omega_0$ = 0.3}. 

\section{Data}
\label{sec:sample}
\subsection{Sample Overview}
In order to investigate the kiloparsec scale properties of star formation and outflows in galaxies at $z\sim$~2.3, we utilise deep adaptive optics assisted near infrared integral field observations from the SINS/zC-SINF AO Survey \citep{NMFS18a}. All of our targets were observed in the K band (1.95-2.45$\mu$m) with the Spectrograph for INtegral Field Observations in the Near Infrared (SINFONI; \citealt{Eisenhauer03, Bonnet04}) on the Very Large Telescope (VLT). The K band observations span a rest-frame wavelength range of at least 6450-6950\AA\ across the full redshift range of our sample. We use the \Ha\ emission line at 6563\AA as a tracer of both star formation and ionized gas outflows (see Sections \ref{subsec:sinfoni_fitting} and \ref{sec:broad_emission}). In this section we present an overview of the SINS/zC-SINF AO sample, and refer the reader to the indicated survey papers for further information.

The SINS/zC-SINF AO Survey targeted 36 star forming galaxies at \mbox{$z\sim$~1.5~--~2.5} with SINFONI. The 36 galaxies were drawn from the Spectroscopic Imaging survey in the Near-infrared with SINFONI \citep[SINS;][]{NMFS09} and the zCOSMOS-SINFONI survey \citep[zC-SINF;][]{Mancini11}, which together provided seeing limited (0.5-0.6'' resolution) \Ha\ observations for 110 star forming galaxies with stellar masses in the range \mbox{2$\times$10$^9$-2$\times$10$^{11}$ M$_\odot$}. The SINS and \mbox{zC-SINF} parent samples were selected from spectroscopically confirmed subsets of several imaging surveys, using a range of criteria to probe different star forming populations at high redshift (K band and 4.5$\mu$m flux selection, sBzK color selection, and the optical BX/BM criteria). The final targets were required to have a secure optical spectroscopic redshift, with the \Ha\ line falling in a region of the SINFONI H or K band filter relatively free of contamination from OH lines. 17 SINS galaxies and 19 zC-SINF galaxies with suitable AO reference stars were chosen for AO follow up, with some preference given to brighter targets (integrated \Ha\ flux \mbox{$\gtrsim$ 10$^{-16}$ erg s$^{-1}$ cm$^{-2}$}). The full width at half maximum (FWHM) of the point spread function (PSF) for the AO observations ranges from \mbox{0.15-0.25''} (median 0.17'', or $\sim$1.4 kpc physical size), and the pixel scale in the final datacubes is 0.05''. The spectral resolution of the observations, measured from sky lines in the unsubtracted data cubes, is R~=~3530 (FWHM~=~85~\kms; see Appendix B of \citealt{NMFS18a}).

\begin{figure}
\centering
\includegraphics[scale=0.7, clip = True, trim = 0 0 0 0]{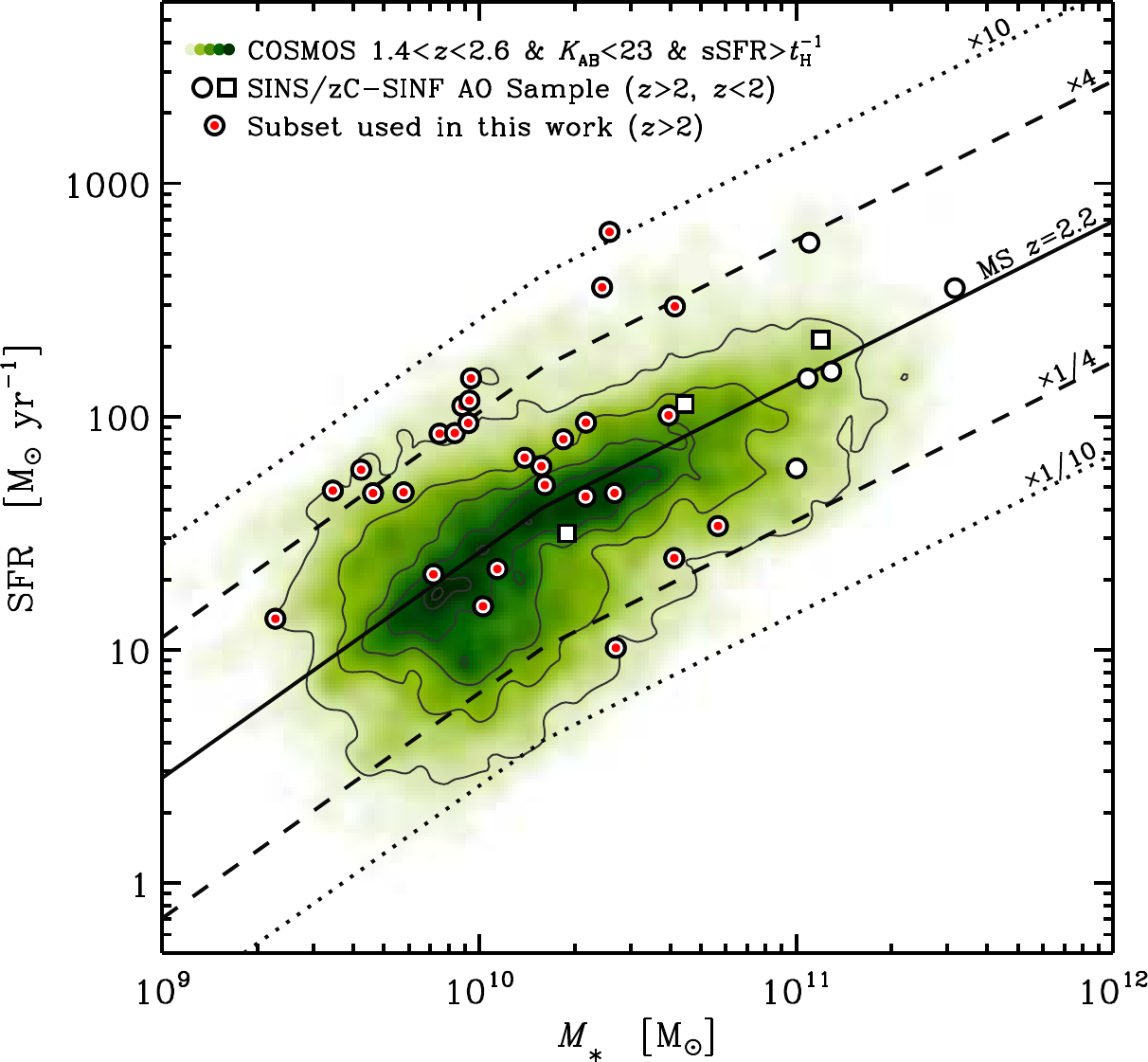}
\caption{Distribution of the SINS/zC-SINF AO sample in the M$_*$-SFR plane. Galaxies at $z>$~2 and $z<$~2 are indicated by circles and squares, respectively. Markers with red centers indicate the subset of galaxies used in this paper (galaxies at $z>$~2 with no evidence for AGN activity). The green contours trace the density distribution of star forming galaxies in the COSMOS field which lie in the redshift range 1.4~$<z<$~2.6 and have K$_{S, AB}$~$<$~23 and inverse specific SFR lower than the Hubble time at the redshift of each object \citep{Ilbert09, Wuyts11}. The solid line indicates the main sequence of star forming galaxies at $z\sim$~2.3 from \citet{Whitaker14}, and the dashed and dotted lines indicate SFRs offset from the main sequence by factors of 4 and 10, respectively. \label{fig:sins_selection}}
\end{figure}

The AO sample is representative of the no-AO SINS~+~zC/SINF parent sample, and covers well the bulk of the $z\sim$~2 star forming galaxy population in the M$_*$-SFR and M$_*$-R$_e$ planes, over 2 orders of magnitude in M$_*$. Figure \ref{fig:sins_selection} shows the M$_*$-SFR distribution of the AO sample (circles and squares), compared to the underlying star forming galaxy population at 1.4~$<~z~<$~2.6 (green contours) and the main sequence of star forming galaxies at $z\sim$~2.3 (\citealt{Whitaker14}, black solid line). Due to the requirement for a pre-existing spectroscopic redshift, redder galaxies are underrepresented and there is some bias towards higher SFRs at low M$_*$. There is no particular size bias in the sample. More detailed discussions of the sample properties can be found in \citet{NMFS18a} and \citet{Tacchella15b}.

\begin{sloppypar}
In this work, we focus on the K band subset of the SINS/zC-SINF AO Survey (33 galaxies at \mbox{z = 2-2.5}). The high resolution AO observations are key for analysing the relationship between star formation properties and outflow properties on kiloparsec scales. Five galaxies were removed from the sample because of evidence for AGN activity (based on mid infrared, X-ray, radio and/or optical emission line indicators) which could significantly contaminate the \Ha\ SFRs. Our final sample consists of 28 galaxies (indicated by the markers with red centers in Figure \ref{fig:sins_selection}) which span 1.4 dex in stellar mass (\mbox{9.4 $\leq$ log(M$_*$/M$_\odot$) $\leq$ 10.8}, median \mbox{log(M$_*$/M$_\odot$) = 10.1}) and 1.8 dex in main sequence offset (\mbox{-0.8 $\leq$ log(SFR/SFR$_{MS}$) $\leq$ 1.0}, median log(SFR/SFR$_{MS}$)~=~0.3).
\end{sloppypar}

\subsection{Global Galaxy Properties}\label{subsec:global_properties}
The global stellar masses, SFRs and visual extinctions ($A_V$) towards all the galaxies in our sample were calculated using standard Spectral Energy Distribution (SED) fitting procedures, described in \citet{NMFS09} and \citet{Mancini11}. In brief, the optical-to-NIR broad-band SEDs (plus the mid-IR 3-8$\mu$m photometry when available) were fit with \citet{BC03} stellar population models, assuming the \citet{Chabrier03} IMF, solar metallicity, the \citet{Calzetti00} reddening law, and constant or expontentially declining star formation histories. The galaxy ages were restricted to be between 50~Myr and the age of the universe at the redshift of each object.

The SINS/zC-SINF targets were drawn from several fields with varying photometric coverage, ranging from four to ten optical-to-NIR bands, supplemented with four-band IRAC \mbox{3-8$\mu$m} data when available (as summarised by \citealt{NMFS09, NMFS11, Mancini11}). More specifically, the SED modelling for the zCOSMOS targets (16/28 galaxies) was based on 10 band optical-to-NIR photometry from Subaru and CFHT \citep{Capak07, McCracken10} and IRAC 3-8$\mu$m photometry \citep{Ilbert09}. For the BX targets (7/28 galaxies) we used the ground based \textit{$U_n$GRJ$K_S$} photometry presented in \citet{Erb06}, supplemented with HST $H_{160}$ measurements from \citet{NMFS11} for three of these objects. One BX target lacked NIR ground-based data but was among the sources imaged in H-band with HST. The SED modelling for the GMASS targets (2/28 galaxies) was based on photometry from HST (
\textit{BVIZ}), VLT (\textit{JH$K_S$}) and IRAC (\mbox{3-8$\mu$m}) \citep{Kurk13}. For the K20 targets (2/28 galaxies) we used the $B$-8$\mu$m photometry from the FIREWORKS catalogue \citep{Wuyts08}, and for the one GDDS target we used the 7 band \mbox{$B$-$K_S$} photometry from the Las Campanas Infrared Survey \citep{Chen02, Abraham04}.

The effective radii, axis ratios and position angles of the galaxies in our sample were derived by fitting a single Sersic component to either the HST F160W (H band) light distribution (when available; see \citealt{Tacchella15b}) or the \Ha\ flux distribution (see Section \ref{subsubsec:linemaps} and \citealt{NMFS18a}). The galaxy inclinations ($i$) were calculated directly from the axis ratios ($q$), using $i$~=~cos$^{-1}$($q$).

\section{Method}
\label{sec:sins_method}

The emission line spectra of galaxies are superpositions of emission associated with different gas components. In normal star forming galaxies at $z\sim$~2, the \Ha\ emission primarily traces star formation in the disk of the galaxy. In integral field data, the centroid of the emission line varies spatially, tracing the underlying orbital motions of the ionized gas. If the galaxy has an outflow, a broader (sometimes blue-shifted) \Ha\ emission component will be superimposed on the emission from the star forming disk. The width and velocity offset of the outflow component are tracers of the outflow velocity \citep[e.g.][]{Rupke05, Veilleux05, Genzel11}, while the total flux in the outflow component is a tracer of the outflowing mass \citep[e.g.][]{Genzel11, Newman12_406690}. 

A first order estimate of the strength and width of the outflow component can be obtained by fitting a single Gaussian component. If there is no outflow component, the measured $\sigma$ will be defined by the core of the line profile, but if there is significant flux in the wings of the emission lines, the best fit Gaussian will broaden to include some of this flux \citep[see also][]{NMFS18a}. Not all of the flux in the wings will be captured by a single Gaussian fit, and therefore the measured $\sigma$ will provide a lower limit on the velocity dispersion of the outflow component ($\sigma_b$). More accurate measurements of the flux and kinematics of the outflow component can be obtained by fitting the \Ha\ line profile as a superposition of two Gaussian components \citep[e.g.][]{Shapiro09, Genzel11, Newman12_global, Newman12_406690, NMFS14, Freeman17, Leung17, NMFS18b}.

We characterise the variation in outflow properties as a function of local physical properties across the SINS/zC-SINF AO galaxies by shifting the spectra of individual spaxels to remove large scale velocity gradients across the galaxies, splitting the spectra into bins in a range of physical properties (SFR, \sfrsd, \smsd, \sfrsd/\smsd, $A_V$, and R/R$_e$), and creating high S/N stacks of the spectra in each bin. We use single component Gaussian fits to investigate the relationships between these six physical quantities and the presence of outflows. We find that the line width is primarily determined by the level of star formation (probed by the SFR and \sfrsd), but that \smsd\ may also play an important role in modulating the presence and properties of outflows. We focus our more detailed investigation on \sfrsd, because it is directly linked to the star formation processes which provide the energy to drive the outflows, and because it is normalised by area, making it easier to consistently compare with measurements at different spatial scales than the SFR. We use two component Gaussian fitting to quantify the relationships between \sfrsd\ and the incidence, velocity and mass loading factor of outflows.

The assumption that the star formation and outflow components are well represented by Gaussian profiles is justified by the central limit theorem, because the star formation component is an average over many \HII\ region spectra, and the outflow component is an average over many outflow spectra \citep[see also][]{NMFS18b}. The validity of the assumption is also confirmed empirically - we find that the emission line profiles of stacks with no evidence for broad emission are well fit by a single Gaussian component \citep[see also][]{Genzel11}, and the line profiles of stacks with clear broad wings are well fit by two Gaussian components, with no evidence for strong asymmetries or blueshifts in the outflow components (see Section \ref{subsec:sigma_fbroad}).

\subsection{Mapping Physical Properties Across Galaxies}
\label{subsec:maps}

\begin{figure*}
\centering
\includegraphics[clip = True, trim = 5 67 8 45, scale = 0.8]{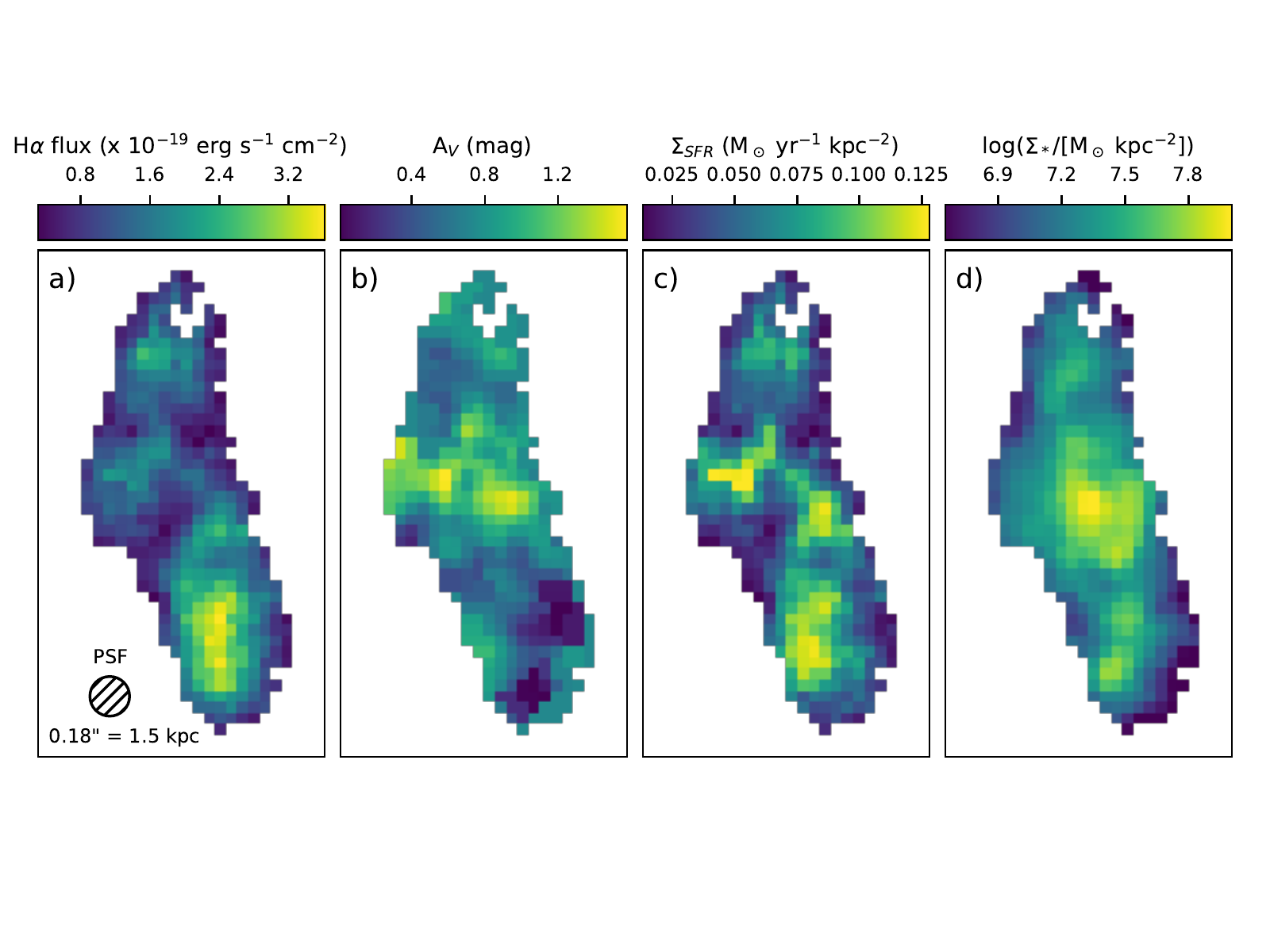}
\caption{Maps of the \Ha\ flux, $A_V$, \sfrsd\ and \smsd\ for ZC405501, at the 0.05'' pixel scale of the reduced data. ZC405501 is at a redshift of $z$~=~2.154, and has a stellar mass of log(M$_*$/M$_\odot$) = 9.9 and a SFR of 85~M$_\odot$~yr$^{-1}$. \label{fig:zc405501_maps}}
\end{figure*}

\subsubsection{\Ha\ Flux Maps}\label{subsubsec:linemaps}
\Ha\ flux, velocity and velocity dispersion ($\sigma$) maps for all of the galaxies in our sample are presented in \citet{NMFS18a}, and the details of the line fitting are described in that paper. Briefly, the \Ha\ line profiles were fit using the IDL emission line fitting code \textsc{linefit} \citep{NMFS09, Davies11}. Before fitting, the datacubes were lightly smoothed by median filtering, both spectrally (over 3 spectral channels) and spatially (over boxes of \mbox{3 $\times$ 3} spatial pixels for 26/28 of the galaxies, and \mbox{5 $\times$ 5} spatial pixels for two large low surface brightness galaxies), to increase the S/N per spaxel. For each spaxel, the continuum was modelled as a straight line through spectral regions adjacent to the line region and free from skyline contamination. The \Ha\ line was modelled as a single Gaussian convolved with the line spread function of SINFONI at the observed wavelength of the line. The spectral channels were inverse variance weighted to prevent strong sky residuals from biasing the fits. The errors on the fit kinematics and \Ha\ flux were calculated from 100 Monte Carlo simulations, where the value of the spectrum at each wavelength was perturbed assuming a Gaussian distribution with a standard deviation given by the input error cube. Spaxels with \mbox{\Ha\ flux/error $<$ 5} are masked in the final maps. Panel a) of Figure \ref{fig:zc405501_maps} shows the \Ha\ flux map for the galaxy ZC405501, at the 0.05'' pixel scale of the reduced data. 

\subsubsection{Galactocentric Distance Maps}
The deprojected radial distance of each spaxel from the kinematic center of each galaxy is calculated using the galaxy position angle and axis ratio. We normalise the deprojected distance maps to the effective radii of the galaxies.

We also create maps of the circularized distance of each spaxel from the centre of each galaxy. The circularized radius r$_{circ}$ is defined as the radius of the circle whose area equals the area of the ellipse with major axis radius $r$ and minor axis radius $r~\times$~$q$ (where $q$ is the axis ratio of the galaxy). The major axis radius is the same as the deprojected radius by definition, and therefore \mbox{r$_{circ}$ = $r_{deproj} \sqrt{q}$}. 

\subsubsection{$A_V$ and $A_{H\alpha}$ Maps}
At $z~\sim$~2, the observed \mbox{HST F438W (B)} - \mbox{F814W (I)} color is a proxy for the rest frame FUV-NUV color, which probes the slope of the ultraviolet continuum and is therefore a good tracer of the dust attenuation. \citet{Tacchella18} investigated the relationship between the FUV-NUV color and the $A_V$, using a grid of model SEDs from \citet{BC03}. The model SEDs cover six different metallicities \mbox{(Z~=~0.0001-0.05)}, three different star formation histories (exponentially increasing, constant, and expontentially declining), and a range of attenuations \mbox{($A_V$~=~0-6)} and redshifts \mbox{($z$~=~2-2.5)}. They considered galaxy ages between 10~Myr and 3.5~Gyr (the age of the universe at $z~\sim$~2) and adopted the \citet{Calzetti00} reddening law.

\citet{Tacchella18} showed that there is a tight, redshift-dependent correlation between the FUV-NUV color and the $A_V$, and used this correlation to create $A_V$ maps for the 10 SINS/zC-SINF AO galaxies with available HST F438W and F814W imaging. 6/10 of these galaxies are included in our sample. The HST maps have similar angular resolution to the AO \Ha\ maps, and were resampled to the same pixel grid. The $A_V$ map for ZC405501 is shown in panel b) of Figure \ref{fig:zc405501_maps}. 

For the 22/28 galaxies without available HST F438W and F814W imaging, it is not possible to directly measure the $A_V$ towards each spaxel. Instead, we combine the average radial $A_V$ profiles derived by \citet{Tacchella18} with the global $A_V$ values measured from the SED fitting (described in Section \ref{subsec:global_properties}) to estimate the $A_V$ towards each spaxel. \citet{Tacchella18} used their 10 $A_V$ maps to create average $A_V$ profiles (in terms of the absolute circularized radius $r_{circ,kpc}$), in two bins of stellar mass. The lower mass (\mbox{log(M$_*$/M$_\odot$) $<$ 11}) bin is comprised entirely of the six galaxies from our sample with $A_V$ maps. For each of the 22 galaxies in our sample without $A_V$ maps, we adopt the average $A_V$ profile of the \mbox{log(M$_*$/M$_\odot$) $<$ 11} galaxies from \citet{Tacchella18}, and scale the profile by a constant factor so that the $A_V$ averaged across the individual spaxels matches the global $A_V$ measured for the galaxy.

There is growing evidence that high redshift star forming galaxies display negative radial attenuation gradients \citep[see also e.g.][]{Wuyts12, Hemmati15, Nelson16, Liu17}. Therefore, the radial dust corrections that we have applied should produce more accurate \Ha\ luminosities than global dust corrections. However, a radial attenuation profile cannot account for azimuthal $A_V$ variations driven by the clumpy distribution of star formation in high redshift galaxies. Furthermore, high mass galaxies are observed to have steeper $A_V$ gradients than low mass galaxies. The six galaxies from which the average profile is constructed are biased towards the high mass end of our sample (\mbox{10.0 $\leq$ log(M$_*$/M$_\odot$) $\leq$ 10.7}), and therefore the average $A_V$ profile is likely to be steeper than the intrinsic $A_V$ profiles of the lower mass galaxies. Better constraints on the $A_V$ towards individual spaxels of our galaxies will be crucial for obtaining more accurate \Ha\ luminosities in the future.

The spaxel $A_V$ values were converted to $A_{H\alpha}$ values using Equation 5 of \citet{Tacchella18}, which assumes that the stellar extinction follows the \citet{Calzetti00} curve and that the nebular extinction follows the \citet{Cardelli89} curve. We assume that the ratio of the E(B-V) for the stellar continuum to the E(B-V) for the nebular emission lines ($f_{\rm */neb}$) is 0.7 (rather than $f_{\rm */neb}$~=~0.44 which is adopted in the local universe; \citealt{Calzetti00}). This choice is motivated by the results of \citet{Tacchella18}, who showed that adopting $f_{\rm */neb}$~=~0.7 produces better agreement between the UV and \Ha\ SFRs than $f_{\rm */neb}$~=~0.44, and \citet{Kashino13}, who measured $f_{\rm */neb}$~=~0.7-0.8 for star forming galaxies at 1.4~$<z<$~1.7 in COSMOS. However, our main conclusions do not change if we adopt $f_{\rm */neb}$~=~0.44.

\subsubsection{SFR and \sfrsd\ Maps}\label{subsubsec:sfr_maps}
\begin{sloppypar}
The SFR and \sfrsd\ maps were derived from the \Ha\ flux maps as follows. The \Ha\ fluxes were corrected for extinction using the calculated $A_{H\alpha}$ values, and converted to luminosities using the galaxy redshifts. The spaxel \Ha\ luminosities were converted to SFRs using the \citet{Kennicutt98} calibration, adjusted to the \citet{Chabrier03} initial mass function (\mbox{SFR [M$_\odot$ yr$^{-1}$] = $L_{\rm H\alpha}$/[2.1$\times$10$^{41}$ erg s$^{-1}$}]). Finally, the SFRs were divided by the deprojected area of each pixel on the sky (accounting for galaxy inclination) to obtain \sfrsd\ in units of \myrkpc. The \sfrsd\ map for ZC405501 is shown in panel c) of Figure \ref{fig:zc405501_maps}.
\end{sloppypar}

\subsubsection{\smsd\ Maps}
\citet{Tacchella15} presented resolved stellar mass maps for 24/28 of the galaxies in our sample. They used HST \mbox{F110W (J) - F160W (H)} color maps to derive pixel mass to light ratios, assuming the \citet{Chabrier03} initial mass function, and multiplied the derived mass to light ratios by the pixel H band luminosities to obtain the stellar mass in each pixel. The stellar mass maps can be directly converted to \smsd\ maps by dividing by the deprojected area of each pixel on the sky. The \smsd\ map for ZC405501 is shown in panel d) of Figure \ref{fig:zc405501_maps}. 

The remaining 4/28 galaxies lack the F110W and/or F160W imaging required to calculate pixel mass to light ratios, and are not used in any analysis requiring \smsd\ estimates. These four galaxies have intermediate stellar masses (\mbox{10.0 $\leq$ log(M$_*$/M$_\odot$) $\leq$ 10.4}) and a range of SFRs (\mbox{10-620~M$_\odot$ yr$^{-1}$}), and we show that the missing resolved stellar mass information is unlikely to bias our results (see Section \ref{sec:broad_emission}).

\subsection{Stacking}\label{subsec:sinfoni_stack_creation}
\subsubsection{Creating Stacks}
We stack the spectra of individual spaxels from the original (unsmoothed) datacubes in bins of the physical properties described above to create high S/N \mbox{(\Ha\ S/N~=~60-170)} stacks which can be used to study the relationships between those physical properties and outflow properties. Our datacubes contain a total of 9510 spaxels which have robustly measured kinematics and are therefore suitable for this analysis (see Section \ref{subsubsec:linemaps}).

The spectrum of each spaxel is shifted and re-sampled so that the centroid of the \Ha\ line is at zero velocity and the spectral channels have a width of 30~\kms\ (which is close to the initial velocity sampling of \mbox{$\sim$30-40~\kms}). The spaxels are divided into bins based on the physical property of interest. Each spaxel is assigned a single weight which is applied to all spectral channels. The weight is given by the inverse of the root mean square (rms) value of the line free channels (more than 40\AA\ from the center of the \Ha\ line and more than 30\AA\ from the center of the \SII$\lambda \lambda$6716,6731 doublet). From the spaxels in each bin, a stack is created by computing the weighted average of the spectra, subtracting a constant continuum value (given by the median value of the line free channels), and normalizing to a maximum value of 1. We also create `unweighted' stacks (where all spaxels are given a weighting of 1), but find that the choice of weighting scheme does not impact our main conclusions.

The errors on the stacked spectra are calculated using bootstrapping. For each bin, we randomly draw half of the spaxels and stack them, and repeat the process 100 times so that we have 100 bootstrap stacks. The error on each spectral channel of the final stack is then given by the standard deviation of the values of the 100 bootstrap stacks for that channel. 

\subsubsection{Quantifying Outflow Properties}\label{subsec:sinfoni_fitting}
We quantify the strength and properties of the outflow component in each stack by fitting each of the \Ha\ and \NII\ lines with either one Gaussian component, or a superposition of two Gaussian components - a narrow component tracing the gas in the star forming regions, and a broader component tracing the outflows. We stress that although we sometimes refer to the outflow component as the `broad' component, the velocity dispersion of this component is $\lesssim$~300~\kms\ and it is therefore much narrower than the broad components associated with e.g. AGN driven outflows or broad line regions.

When an outflow component is present, the base of the \Ha\ line can become blended with the bases of the \NII$\lambda$6548 and \NII$\lambda$6584 lines which lie at separations of \mbox{-670 \kms} and \mbox{+940 \kms}, respectively. It is therefore necessary to fit the two \NII\ lines simultaneously with \Ha. In our fitting, we assume that all three lines have the same kinematics; i.e. for each Gaussian component all lines have a common velocity shift and velocity dispersion. 

The emission line fitting is performed using the Python module \textsc{emcee} \citep{ForemanMackey13}, an Affine Invariant Markov Chain Monte Carlo (MCMC) Ensemble Sampler. The MCMC provides the posterior probability distribution for each of the parameters, as well as the joint posterior probability distribution for each pair of parameters, allowing us to ensure that the parameters of interest (namely the \Ha\ amplitude(s) and the velocity dispersion(s)) are well constrained and not degenerate with each other or other parameters in the fit. This is particularly important when fitting two kinematic components as multiple solutions may be possible (see also discussion in \citealt{Freeman17}).

We adopt uniform priors on all model parameters, limiting them to ranges which are physically reasonable. Specifically, the parameters are limited to the following ranges: velocity shifts \mbox{$\abs{\Delta v}$ $<$ 200 \kms}, velocity dispersion of the narrow component ($\sigma_n$) greater than 36~\kms\ (the spectral resolution of the data), velocity dispersion of the outflow component ($\sigma_b$) greater than 150~\kms\ (to minimise contamination from beam smearing; see Appendix \ref{sec:beam_smearing}), and line amplitudes greater than zero. The ratio of the amplitudes of the two \NII\ lines (\NII$\lambda$6584/\NII$\lambda$6548) in each component is fixed to 3 (the theoretical value set by quantum mechanics), but the \NIIHa\ ratio is left as a free parameter. 

We run the MCMC with 400 walkers, 400 burn-in steps and 1200 run steps, based on the calculated auto-correlation times for the fit parameters (110-120 steps). For each parameter, the best fit value is defined as the peak of the posterior probability distribution, and the 68 per cent confidence interval is defined as the smallest interval containing 68 per cent of the probability. 

\begin{figure*}
\centering
\includegraphics[scale=1.07, clip = True, trim = 20 65 20 10]{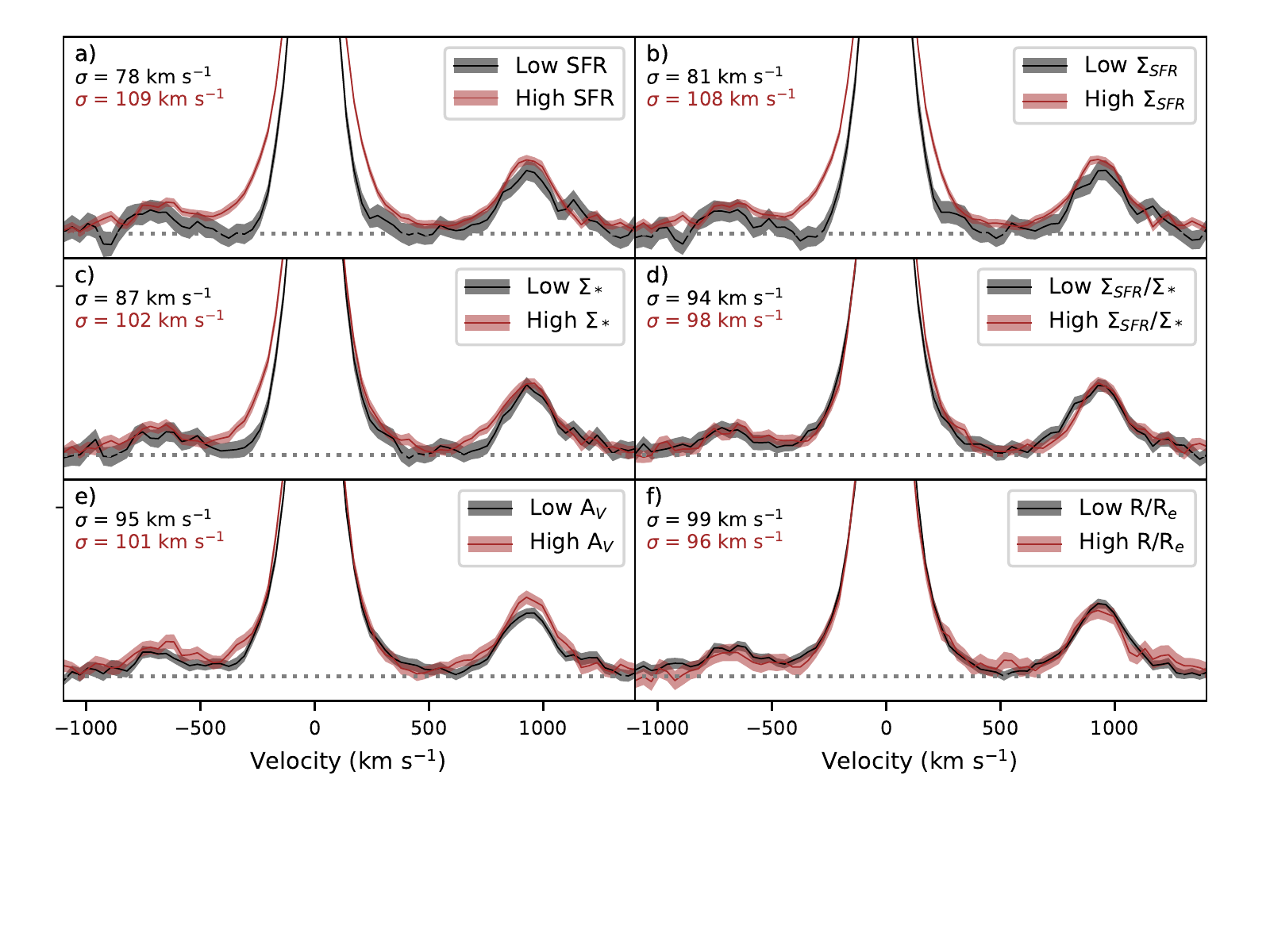}
\caption{Spectra stacked in 2 bins of SFR (panel a), \sfrsd\ (b), \smsd\ (c), \sfrsd/\smsd\ (d), $A_V$ (e) and R/$R_e$ (f). The bins are divided by the median value of each property, listed in Table \ref{tab:sample_halves}. All stacks are normalised to the same \Ha\ amplitude. The filled regions represent the 1$\sigma$ errors derived using bootstrapping. We fit a single Gaussian to the \Ha\ and \NII\ lines in each stack, and find that the line widths are primarily driven by the level of star formation (probed by the SFR and \sfrsd), although \smsd\ may also play a significant role.}\label{fig:sample_halves}
\end{figure*}

\subsubsection{Calculating the Average \sfrsd\ for Each Stack}\label{subsubsec:broad_contam_correction}
To first order, the average \sfrsd\ for each stack is simply given by the average of the \sfrsd\ values of the spaxels that went into the stack. However, the \sfrsd\ values measured for individual spaxels are somewhat biased. For the vast majority of spaxels it is not possible to robustly separate the star formation and outflow components, even when \mbox{S/N(\Ha)~$>$~5}. Therefore, we calculate \sfrsd\ using the single Gaussian \Ha\ flux, which is robust and independent of S/N for S/N(\Ha)~$>$~5, regardless of the strength and width of the outflow component. However, in spaxels with significant outflow components, the single Gaussian \Ha\ flux will be contaminated by emission from the outflow component and therefore \sfrsd\ will be over-estimated relative to other spaxels (assuming that the dust extinction is not preferentially underestimated in regions with outflows).

We account for the contribution of the outflow component a-posteriori by calculating how much (on average) the measured \sfrsd\ values of the spaxels in each stack are over-estimated. For each stack, we fit the \Ha\ and \NII\ lines with one Gaussian component and then with two Gaussian components. We calculate a `correction factor', \Ha(narrow)/\Ha(1~comp). The denominator is the \Ha\ flux from the single component fit, which is the biased value that was used to estimate \sfrsd\ in Section \ref{subsubsec:sfr_maps}. The numerator is the \Ha\ flux of the narrow (star formation) component from the two component fit, which is the value that should be used to calculate the true \sfrsd.

The intrinsic average \sfrsd\ for spaxels in each bin can therefore be calculated from the measured average \sfrsd\ and the correction factor as follows:
\begin{equation}
\Sigma_{\rm SFR} ({\rm intrinsic}) = \Sigma_{\rm SFR} ({\rm measured}) \times \frac{\rm{H}\alpha~({\rm narrow})}{\rm{H}\alpha~({\rm 1~comp})} \label{eqn:sfrsd_correction}
\end{equation}
The correction factors for our stacks range from \mbox{1.1-1.9}. In the following sections, all quoted \sfrsd\ values have been corrected using Equation \ref{eqn:sfrsd_correction}.

\section{Dependence of Line Width on Resolved Physical Properties}
\label{sec:broad_emission}
We begin by investigating how the strength of the broad component relates to different physical properties. We consider four properties which are thought to be linked to star formation driven outflows (SFR in the spaxel, \sfrsd, \smsd\ and \sfrsd/\smsd), and two properties related to processes which could be potential sources of contaminating broad emission ($A_V$ and galactocentric distance). The $A_V$ probes the amount of dust along the line of sight. If there is a large amount of dust present, some of the \Ha\ light may be scattered to different frequencies, inducing artificial broadening of the emission line (see e.g. \citealt{Scarrott91}). The \Ha\ line could also be artificially broadened by unresolved orbital motions (beam smearing). This effect is particularly prominent in the centers of massive galaxies where velocity gradients are the largest \citep[e.g.][]{Epinat10, Davies11, Newman13}. 

Figure \ref{fig:sample_halves} shows how the shape of the \Ha\ line varies as a function of SFR, \sfrsd, \smsd, \sfrsd/\smsd, $A_V$ and R/R$_e$. These properties are known to correlate with one another, so we look for the property which shows the most pronounced correlation with the strength of the broad component. For each property, the spaxels are divided into two bins (above and below the median value listed in Table \ref{tab:sample_halves}), and a stack is created for each bin as described in Section \ref{subsec:sinfoni_stack_creation}. In each panel, the stack of spaxels above the median is shown in red, and the stack of spaxels below the median is shown in black. For each stack, the filled region indicates the 1$\sigma$ error region. 

We fit each of the \Ha\ and \NII\ lines in each stack with a single Gaussian, as described in Section \ref{subsec:sinfoni_fitting}. The velocity dispersions of the best fit Gaussians for all the stacks are listed in Figure \ref{fig:sample_halves}. The differences between the $\sigma$ values measured for the above and below median stacks for each property are listed in Table \ref{tab:sample_halves}. All velocity dispersions quoted in this paper have been corrected for the spectral resolution (FWHM~=~85~\kms). The formal uncertainties on the fit $\sigma$ values are very small ($\sim$~1~\kms), but do not account for the fact that a single component Gaussian model is sometimes not a good representation of the data (see also discussion at the beginning of Section \ref{sec:sins_method}), and therefore the uncertainties are not very meaninfgul.

\begin{table}[]
\centering
\begin{tabular}{c|c|c}
Property & Median Value & $\Delta \sigma$ (\kms) \\ \hline
SFR & 0.08 M$_\odot$ yr$^{-1}$ & 31 \\
$\Sigma_{SFR}$ & 0.19 \myrkpc & 27 \\
$\Sigma_*$ & 10$^{7.78}$ M$_\odot$ kpc$^{-2}$ & 15 \\
$\Sigma_{SFR}/\Sigma_*$ & 10$^{-8.4}$ yr$^{-1}$ & 4 \\
$A_V$     &  0.93 mag & 6 \\
R/$R_e$    &  1.22 & -3 \\ \hline
\end{tabular}
\caption{Quantities related to the pairs of stacked spectra in Figure \ref{fig:sample_halves}. The spaxels were divided into two bins for each of the properties in column 1, with one bin above and one bin below the median values in column 2. Column 3 lists the difference between the single component velocity dispersions fit to the above and below median stacks for each property.}
\label{tab:sample_halves}
\end{table} 

Figure \ref{fig:sample_halves} shows that the shape of the emission line profiles is most strongly dependent on the level of star formation, probed by the SFR and \sfrsd\ (panels a and b). The high SFR stack shows a clear excess of flux at high velocities with respect to the low SFR stack, and the same is true for the \sfrsd\ stacks. The measured velocity dispersions for the above and below median SFR stacks are $\sigma$~=~109~\kms\ and 78~\kms, respectively, corresponding to $\Delta \sigma$~=~31~\kms. Similarly, the measured velocity dispersions for the above and below median \sfrsd\ stacks are $\sigma$ = 108~\kms\ and 81~\kms, respectively, corresponding to $\Delta \sigma$~=~27~\kms. The distribution of spaxels between the above and below median bins does not change significantly when dividing based on SFR or \sfrsd, because the SFRs of the spaxels in our sample vary by a factor of 50, but the deprojected areas of the pixels on the sky only vary by a factor of four. The $\Delta \sigma$ values measured for the SFR and \sfrsd\ stacks are significantly larger than the $\Delta \sigma$ values measured for the \smsd, \sfrsd/\smsd, $A_V$ and R/R$_e$ stacks, suggesting that there is a direct link between the level of star formation activity and the strength of the broad emission component. This is consistent with predictions that the incidence and velocity of star formation driven outflows should increase with \sfrsd\ \citep[e.g.][]{Thompson05, Ostriker11, Faucher13, Hayward17}. 

The shape of the emission line profiles also varies with \smsd\ (panel c of Figure \ref{fig:sample_halves}). The high \smsd\ stack shows an excess of flux at high velocities compared to the low \smsd\ stack, and we measure $\Delta \sigma$~=~15~\kms\ (48 per cent of the $\Delta \sigma$ between the SFR stacks). In contrast, we do not observe any significant trends in emission line width as a function of \sfrsd/\smsd, $A_V$ or R/R$_e$ (panels d, e and f of Figure \ref{fig:sample_halves}), and the measured $\Delta \sigma$ values are low (4, 6 and -3 \kms, respectively). 

The absence of a strong correlation between $A_V$ and $\sigma$ indicates that scattered light cannot be the dominant source of broad emission in our sample. The absence of a strong correlation between galactocentric distance and $\sigma$ indicates both that the broad component cannot originate primarily from beam smearing, and that the outflows are launched from a range of radii and are therefore unlikely to be AGN driven (as expected, because AGN hosts were explicity excluded from the sample). The impact of beam smearing on our results is further explored with the aid of dynamical models in Appendix \ref{sec:beam_smearing}.

We note that the \smsd\ and \sfrsd/\smsd\ stacks include spectra from the 24/28 galaxies which have HST F110W and F160W imaging, whereas the SFR, \sfrsd, A$_V$ and R/R$_e$ stacks contain spectra from all 28 galaxies. To assess whether our results are biased by using different sets of galaxies in different stacks, we create SFR, \sfrsd, A$_V$ and R/R$_e$ stacks using the same 24/28 galaxies included in the \smsd\ stacks, and repeat the single component Gaussian fits. The fit $\sigma$ values change by a maximum of 3~\kms, indicating that our results are unlikely to be biased by the missing \smsd\ measurements. 

This analysis assumes that any variation in the measured $\sigma$ is primarily attributable to variations in the strength and/or width of the outflow component, and is not driven by changes in the velocity dispersion of the narrow component, $\sigma_n$. The $\sigma_n$ may be correlated with the level of star formation because energy injection from stellar feedback may contribute to increasing the turbulent pressure in the disk. Furthermore, \smsd\ may be correlated with the gas surface density $\Sigma_g$, which regulates $\sigma_n$ through the Toomre $Q$ parameter \citep[e.g.][]{NMFS06, Genzel11, Krumholz18}. However, the pairs of stacks with the largest $\Delta \sigma$ values in Table \ref{tab:sample_halves} also have the most clearly visible differences in the strength of the broad wings in Figure \ref{fig:sample_halves}, which verifies that the $\Delta \sigma$ is tracing real differences in the strength and width of the broad component, with only a minor secondary dependence on $\sigma_n$ variations.

We conclude that the line width is primarily driven by the level of star formation (probed by the SFR and \sfrsd), but that \smsd\ may also play a significant role in modulating the shapes of the line profiles (consistent with the results of \citealt{Newman12_global}). In this paper we choose to focus on \sfrsd, because it is directly linked to the star formation processes which provide the energy to drive the outflows, and because it is normalised by area, making it easier to consistently compare with measurements at different spatial scales than the SFR.

\begin{figure*}
\centering
\includegraphics[scale=0.6, clip = True, trim = 25 80 30 0]{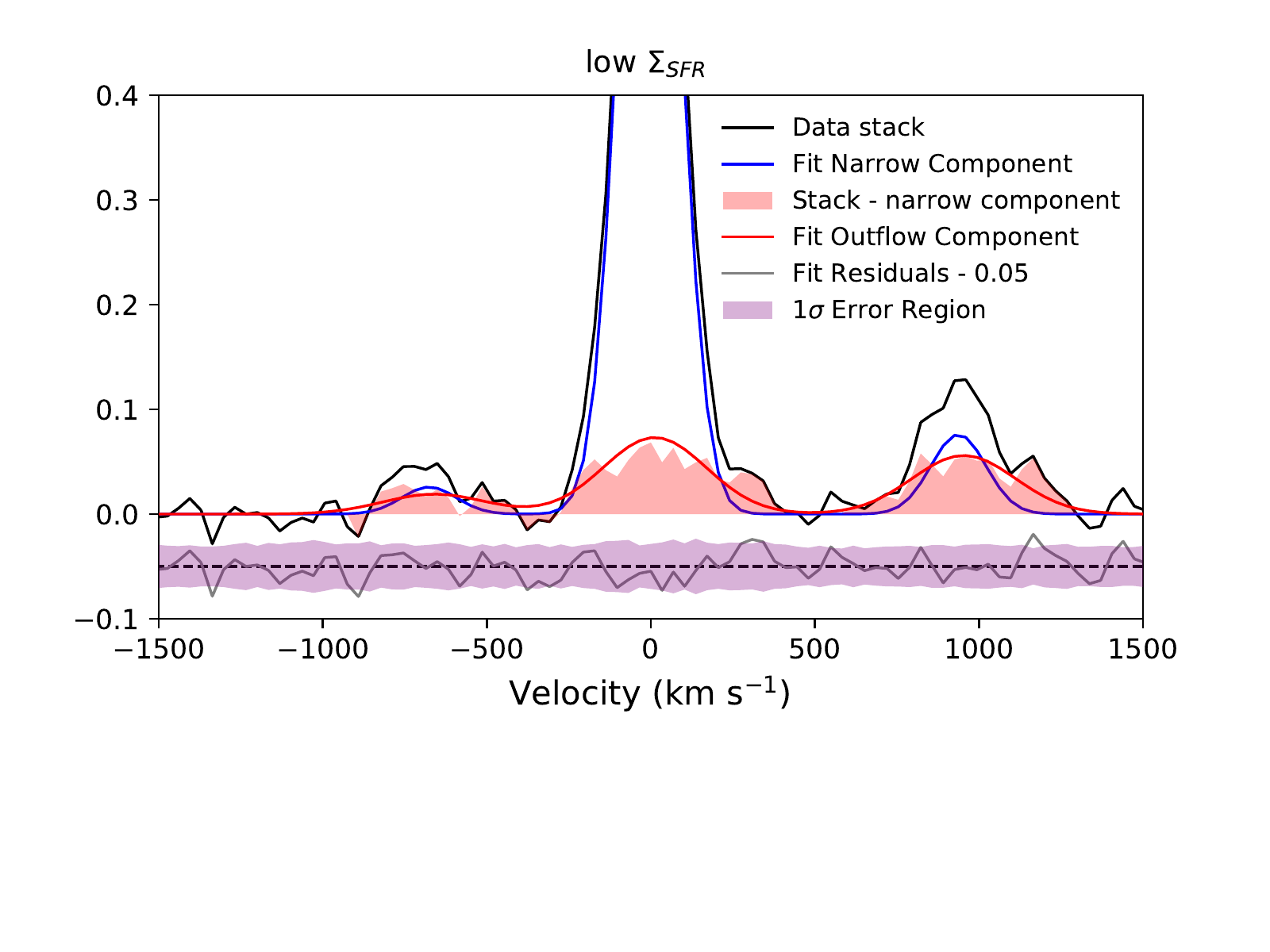}\includegraphics[scale=0.6, clip = True, trim = 25 80 30 0]{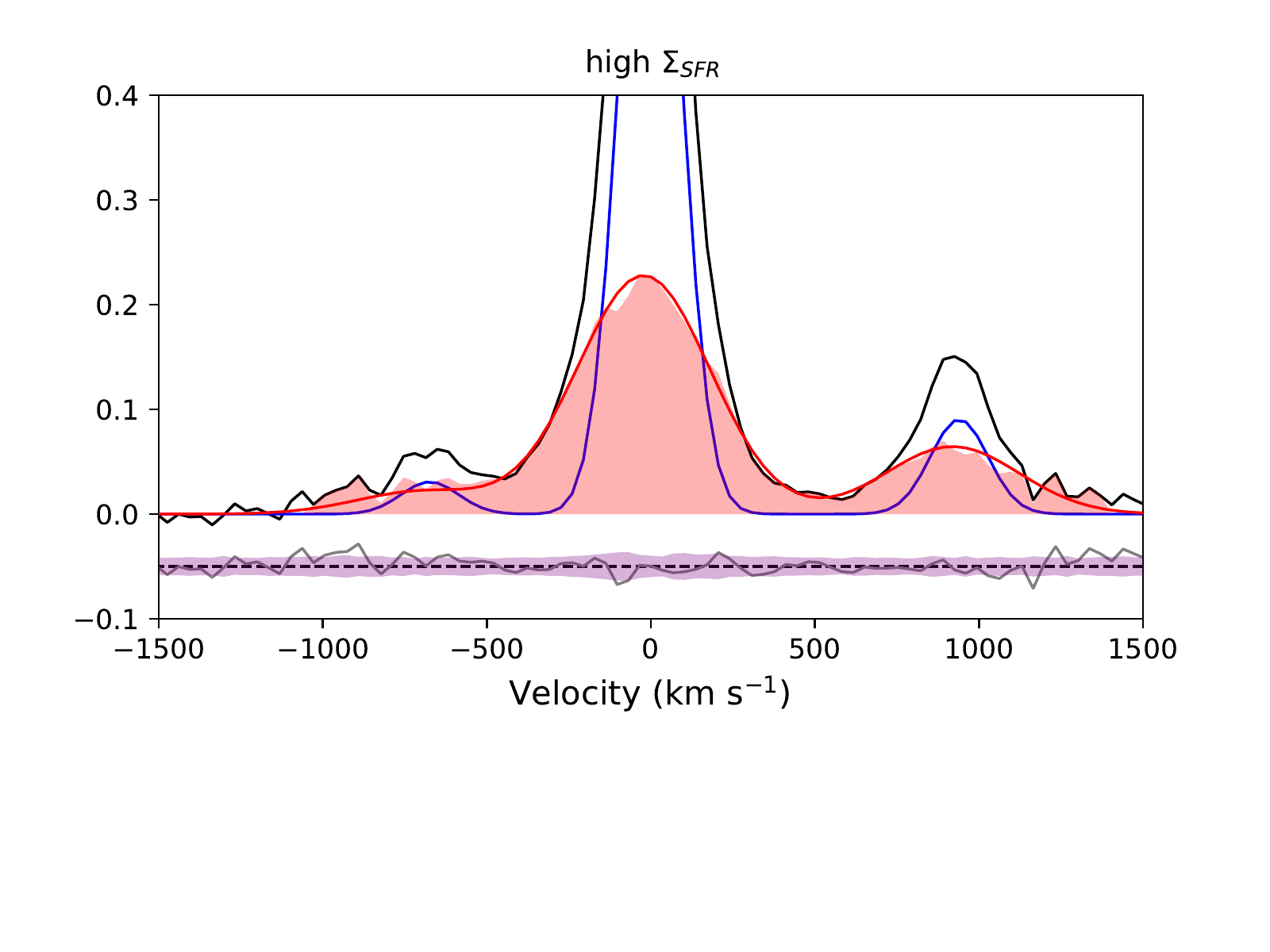} \\
\includegraphics[scale=0.58, clip = True, trim = 20 0 10 0]{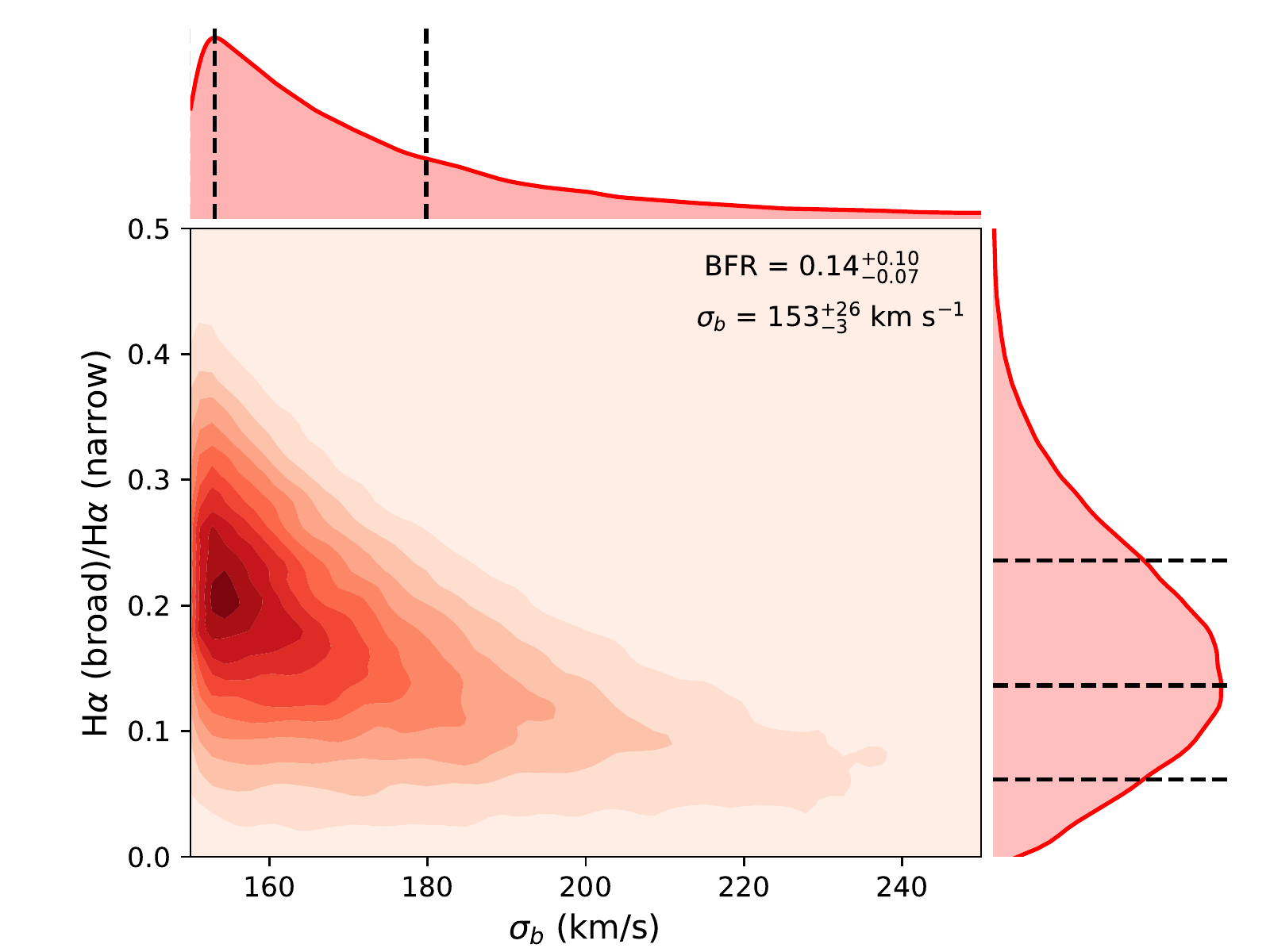}\includegraphics[scale=0.58, clip = True, trim = 20 0 10 0]{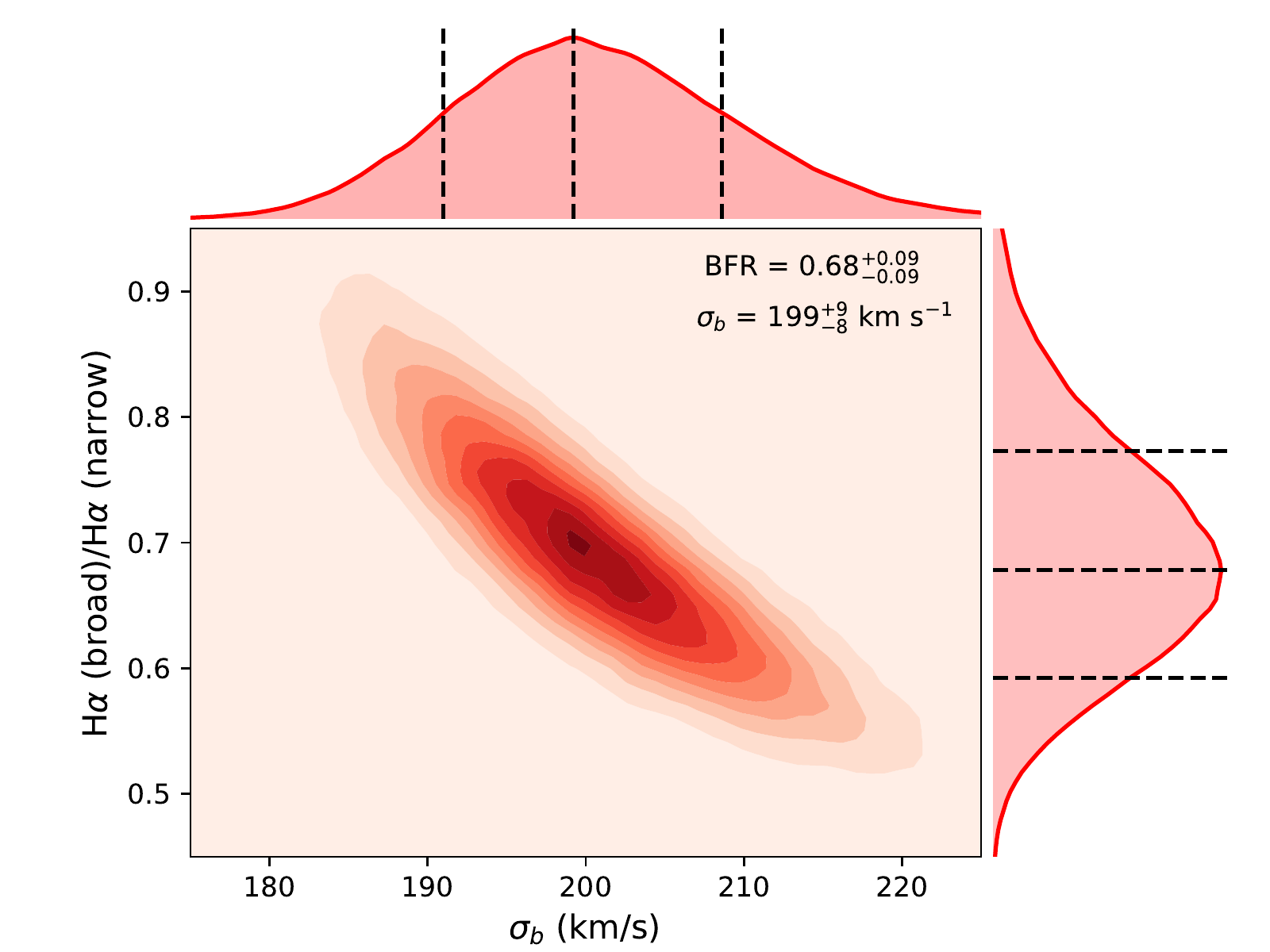}
\caption{Two component Gaussian fits to the above and below median \sfrsd\ stacks (top), and corresponding posterior and joint posterior distributions for the \Ha\ broad-to-narrow flux ratio and the velocity dispersion of the broad component $\sigma_b$ (bottom). In the top panels the stack spectra are shown in black, the best fit narrow components are shown in blue, the residuals after subtracting the fit narrow components from the stack spectra are shown by the pale red shaded regions, and the best fit outflow components are shown with red solid curves. The overall fit residuals are shown in grey and the 1$\sigma$ error regions are shown in purple, and both are artificially offset below zero for clarity.} 
\label{fig:sample_halves_fits}
\end{figure*}

\section{Outflow Properties as a Function of \sfrsd}\label{subsec:outflow_properties}
We further quantify the difference between the line profiles of the low and high \sfrsd\ stacks by fitting the \NII\ and \Ha\ lines as superpositions of two Gaussian components (as described in Section \ref{subsec:sinfoni_fitting}) and comparing the fit parameters. In particular, we focus on parameters which are used to calculate the outflow velocity and mass loading factor (see Sections \ref{subsec:vout} and \ref{subsec:mass_loading}): the ratio of the \Ha\ flux in the broad component to the \Ha\ flux in the narrow component (\Ha\ broad flux ratio or BFR), and the velocity dispersion of the broad component ($\sigma_b$).

\begin{table*}[]
\centering
\begin{tabular}{c|c|c|c|c|c|c|c}
1) & 2) & 3) & 4) & 5) & 6) & 7) & 8) \\ 
$\Sigma_{SFR}~(M_\odot~\rm{yr}^{-1}~\rm{kpc}^{-2})$ & BFR & $\sigma_b$ (\kms) & $\Delta v$ (\kms) & \vout\ & $\eta$ & $\dot{E}_{out}/(10^{-3}~L_{bol})$ & $\dot{p}_{out}/(L_{bol}/c)$ \\ \hline
\multicolumn{8}{c}{Above and below median stacks} \\ \hline
0.11$\pm$0.07  &  0.14$^{+0.10}_{-0.07}$  &  153$^{+26}_{-3}$  &  9$^{+7}_{-14}$  &  296$^{+54}_{-15}$  &  0.05$\pm$0.03  &  0.05$\pm$0.03  &  0.09$\pm$0.04 \\

0.41$^{+0.20}_{-0.23}$  &  0.68$\pm$0.09  &  199$^{+9}_{-8}$  &  -20$\pm$6  &  418$^{+21}_{-15}$  &  0.31$\pm$0.03  &  0.46$\pm$0.03  &  0.64$\pm$0.04 \\ \hline

\multicolumn{8}{c}{5 bin stacks} \\ \hline
0.25$\pm$0.04 &  0.31$^{+0.18}_{-0.09}$  &  180$^{+40}_{-15}$  &  -13$\pm$17  &  373$^{+82}_{-35}$  &  0.13$^{+0.05}_{-0.03}$  &  0.18$\pm$0.05  &  0.28$^{+0.05}_{-0.07}$ \\

0.31$\pm$0.06  &  0.73$^{+0.33}_{-0.16}$  &  166$^{+21}_{-7}$  &  -22$^{+8}_{-10}$  &  355$^{+43}_{-18}$  &  0.29$^{+0.09}_{-0.05}$  &  0.35$\pm$0.05  &  0.54$^{+0.10}_{-0.07}$ \\

0.45$\pm$0.07  &  0.92$^{+0.28}_{-0.11}$  &  183$^{+13}_{-8}$  &  -16$\pm$8  &  384$^{+27}_{-18}$  &  0.40$^{+0.07}_{-0.05}$  &  0.51$\pm$0.06  &  0.77$\pm$0.09 \\

0.80$\pm$0.09  &  0.62$^{+0.18}_{-0.12}$  &  240$^{+27}_{-18}$  &  -24$^{+13}_{-15}$  &  504$^{+56}_{-39}$  &  0.33$^{+0.07}_{-0.04}$  &  0.71$^{+0.15}_{-0.07}$  &  0.85$\pm$0.10 \\

1.44$^{+0.12}_{-0.52}$  &  0.79$^{+0.16}_{-0.11}$  &  263$^{+20}_{-12}$  &  -67$^{+13}_{-16}$  &  595$^{+42}_{-30}$  &  0.51$^{+0.07}_{-0.05}$  &  1.51$^{+0.20}_{-0.12}$  &  1.51$\pm$0.13 \\ \hline
\end{tabular}
\caption{Weighted average \sfrsd, fit parameters and derived outflow properties for each of the \sfrsd\ stacks. \mbox{1) Weighted} average \sfrsd\ of the spaxels in the stack, corrected for broad emission contamination using Equation \ref{eqn:sfrsd_correction}. 2) \Ha\ broad-to-narrow flux ratio. 3) Velocity dispersion of the broad component. \mbox{4) Velocity} offset between the centroids of the broad and narrow components. 5) Outflow velocity, calculated using \mbox{\vout\ = $\Delta$v - 2$\sigma_b$}. \mbox{6) Mass} loading factor ($\dot{M}_{\rm out}$/SFR), assuming $n_e$~=~380~cm$^{-3}$ and $R_{\rm out}$~=~1.7~kpc. 7) Ratio of the measured energy outflow rate to the predicted energy outflow rate for winds driven by energy from supernova explosions. 8) Ratio of the measured momentum outflow rate to the predicted momentum outflow rate for winds driven by radiation pressure from massive stars.}
\label{tab:plot_values}
\end{table*} 

Figure \ref{fig:sample_halves_fits} shows the two component fits to the above and below median \sfrsd\ stacks (top panels), and the corresponding posterior and joint posterior distributions for the BFR and $\sigma_b$ (bottom panels). It is clear that both stacks are well fit by a two component Gaussian. The fit residuals (grey) are generally smaller than the 1$\sigma$ bootstrap errors on the spectra (purple shaded regions). The red shaded outflow components do not exhibit significant asymmetries and are not significantly offset in velocity space from the best fit narrow components, supporting our assumption that both the narrow and broad components are approximately Gaussian in shape.

The high \sfrsd\ stack has a strong outflow component, with \mbox{BFR~=~0.68~$\pm$~0.09} and \mbox{$\sigma_b$~=~199$^{+9}_{-8}$~\kms}. The outflow parameters are well constrained and have approximately Gaussian posterior distributions. In contrast, the MCMC posteriors for the low \sfrsd\ stack reveal that the outflow component contains a small fraction of the total \Ha\ flux and is only marginally detected above the beam smearing limit of 150~\kms. The best fit parameters are \mbox{$\sigma_b$~=~153$^{+26}_{-3}$} \kms\ and BFR~=~0.14$^{+0.10}_{-0.07}$.

To further investigate how the properties of the outflow component vary as a function of \sfrsd, we split the above median \sfrsd\ bin into 5 smaller bins and perform two component Gaussian fitting on each of the resulting stacks. These 5 bins do not have equal numbers of spaxels - rather, they were chosen to span the range of observed \sfrsd\ values with approximately even logarithmic spacing. The weighted average \sfrsd, fit parameters and derived outflow parameters for all stacks are listed in Table \ref{tab:plot_values}. In the following sections we explore trends between \sfrsd\ and outflow properties.

\begin{figure*}
\centering
\includegraphics[scale=0.7, clip = True, trim = 10 10 10 10]{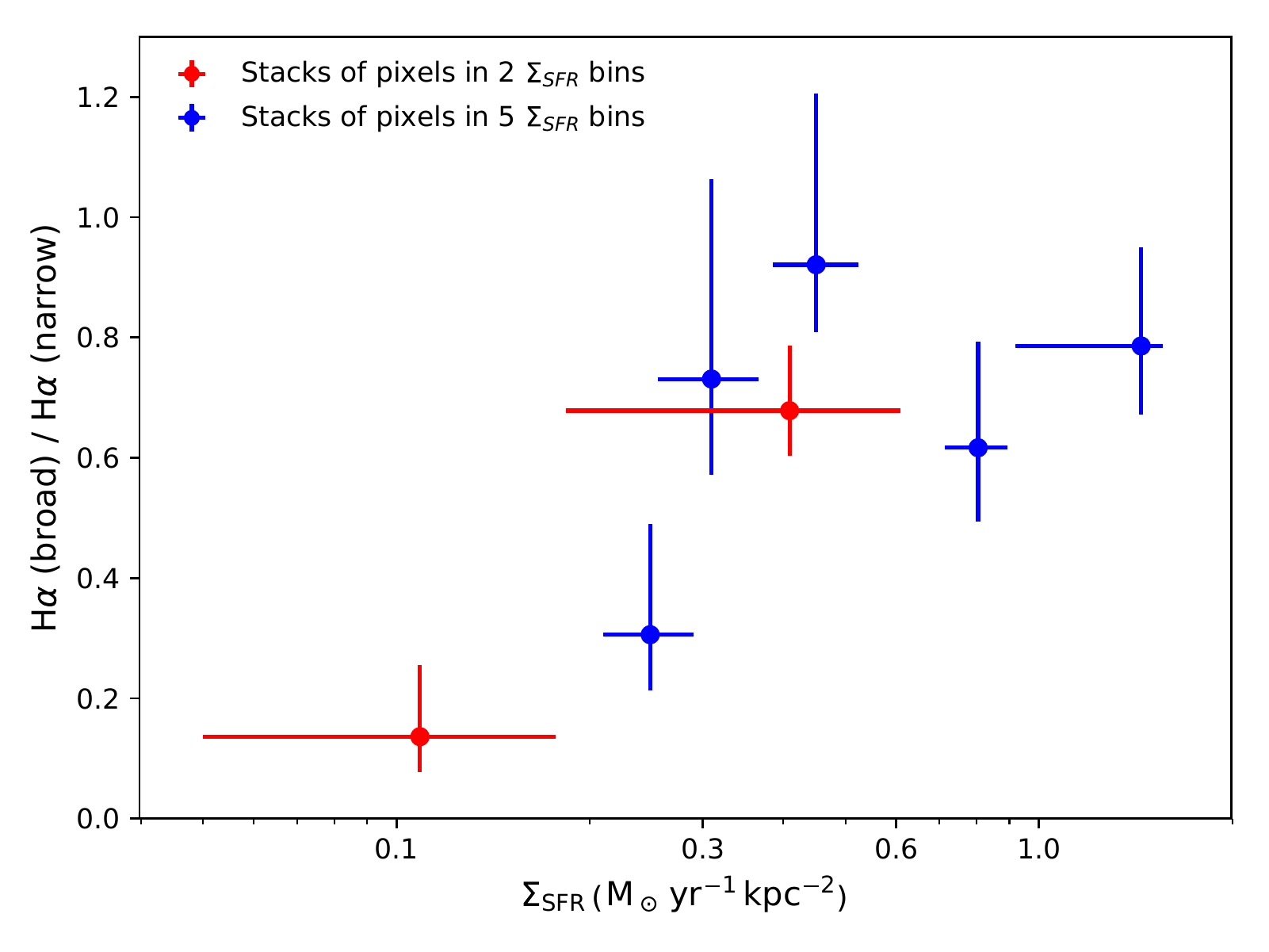}
\caption{Ratio of the \Ha\ flux in the broad component to the \Ha\ flux in the narrow component as a function of \sfrsd. Red symbols represent stacks of spaxels above and below the median \sfrsd, and blue symbols represent stacks of spaxels in five bins of \sfrsd. The error bars on the \sfrsd\ values represent the 16th-84th percentile range of \sfrsd\ values for the spaxels included in each stack, and the error bars on the BFR values represent the 68 per cent confidence interval from the MCMC fitting. The broad component accounts for an average of $\sim$10\% of the total \Ha\ flux at \sfrsd~$<$~0.2~\myrkpc, but increases to $\sim$45\% of the \Ha\ flux at \sfrsd~$>$~0.3~\myrkpc.}
\label{fig:sfrsd_fbroad}
\end{figure*}

\subsection{\Ha\ Broad-to-Narrow Flux Ratio (BFR)}
Figure \ref{fig:sfrsd_fbroad} shows the broad flux ratios measured for the above and below median \sfrsd\ stacks (red circles) and for the five new \sfrsd\ stacks (blue circles). The errors on the \sfrsd\ values represent the 16th-84th percentile range of \sfrsd\ values for the spaxels included in each stack, and the error bars on the BFR values represent the 68 per cent confidence intervals from the MCMC fitting. 

\begin{sloppypar}
The BFR increases from $\sim$0.15 at \mbox{\sfrsd~$<$~0.2~\myrkpc} to $\sim$0.75 at \mbox{\sfrsd~$>$~0.3~\myrkpc}. The factor of 5 increase in the BFR over a small range in \sfrsd\ suggests that outflows are only launched from regions with sufficiently high local \sfrsd.
\end{sloppypar}

\begin{figure*}
\centering
\includegraphics[scale=0.7, clip = True, trim = 10 10 10 10]{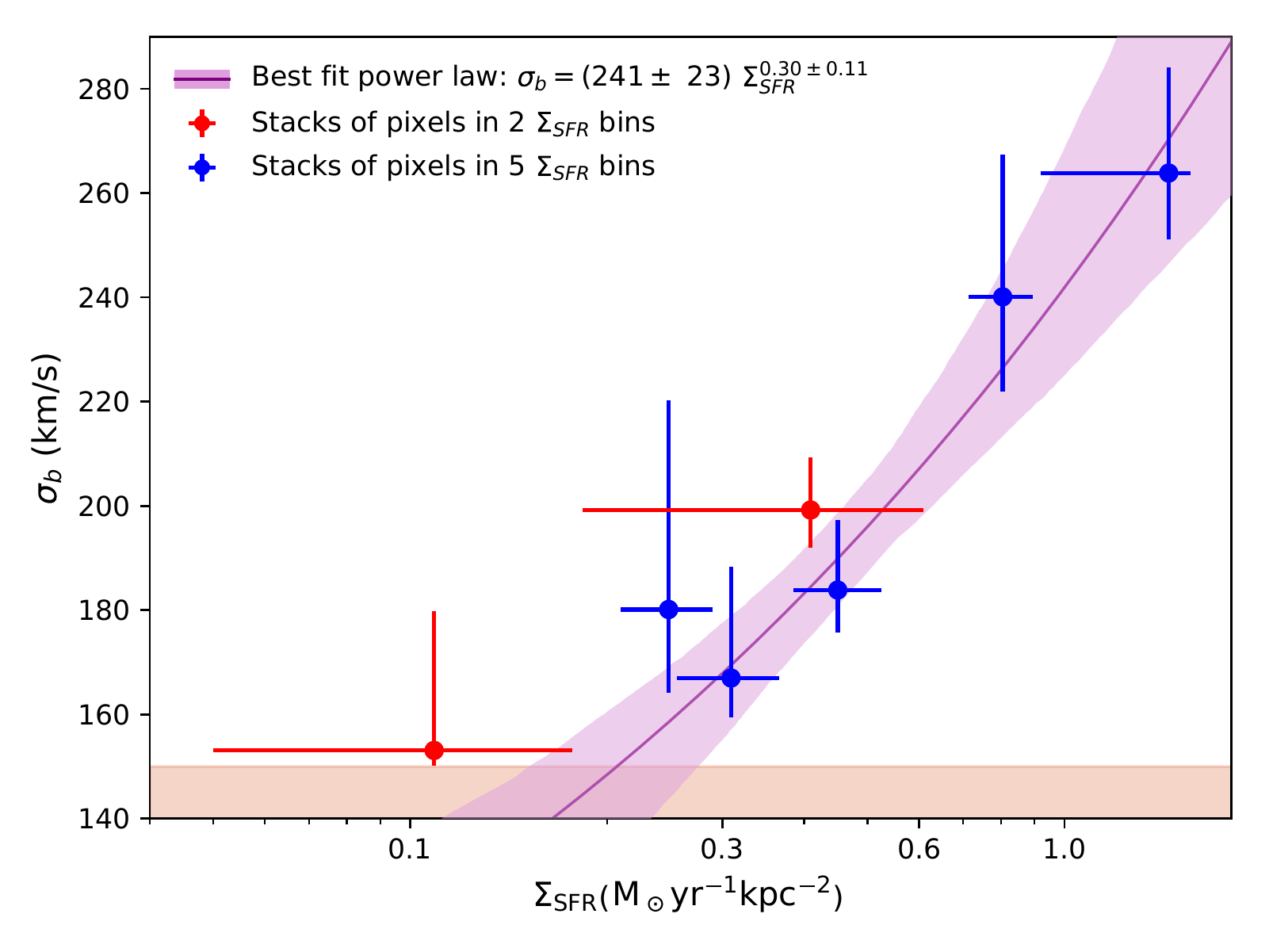} 
\caption{Velocity dispersion of the broad component as a function of \sfrsd. The purple line indicates the best fit power law relation between \sfrsd\ and $\sigma_b$ for our data, and the filled purple region indicates the 1$\sigma$ error around the best fit. The red shading indicates the region at $\sigma_b <$~150~\kms\ which was excluded from our analysis because broad emission components in this region could be dominated by beam smearing and/or stacking artefacts.}
\label{fig:sfrsd_sig}
\end{figure*}

\citet{Newman12_global} analysed the integrated spectra of many of the galaxies in our sample and found that outflows are only driven from galaxies with global \sfrsd\ exceeding 1~\myrkpc\ -- a factor of 3-4 higher than the threshold local \sfrsd\ suggested by our data. If outflows arise from the regions with the highest \sfrsd, then the average \sfrsd\ across a galaxy with outflows should be lower than the \sfrsd\ of the regions actually driving the outflows. The majority of the discrepancy can be explained by the difference between the extinction prescriptions adopted in \citet{Newman12_global} and in this work. In this work, we follow the methodology of \citet{Tacchella18} who assume that the stellar extinction follows the \citet{Calzetti00} curve and the nebular extinction follows the \citet{Cardelli89} curve. We adopt \mbox{$f_{\rm */neb}$~=~0.7}, from which it follows that \mbox{$A_{H\alpha}$ = 0.83 $A_V$}. \citet{Newman12_global} assume that both the stellar and the nebular extinction follow the \citet{Calzetti00} curve, and they adopt \mbox{$f_{\rm */neb}$~=~0.44}, from which it follows that \mbox{$A_{H\alpha}$ = 1.86 $A_V$}. If we adopt the same extinction prescription as \citet{Newman12_global}, the median \sfrsd\ of the spaxels in our sample increases by a factor of 2.2.

Another source of difference is the fact that \citet{Newman12_global} do not account for the contribution of the broad component to the measured \Ha\ fluxes. If we did not apply the correction described in Section \ref{subsubsec:broad_contam_correction}, our \sfrsd\ values would be a factor of 1.5 higher. Therefore, the offset between our local \sfrsd\ threshold and the global \sfrsd\ threshold reported by \citet{Newman12_global} can be fully explained by differences in the adopted extinction prescription, and by the fact that they do not account for the contribution of the broad emission to the measured \Ha\ fluxes.

\subsection{Velocity Dispersion of the Broad Component ($\sigma_b$)}\label{subsec:sigma_fbroad}
Figure \ref{fig:sfrsd_sig} shows how the velocity dispersion of the broad component varies as a function of \sfrsd. The red shading at the bottom of the plot indicates the region at \mbox{$\sigma_b <$~150~\kms} which was excluded from our analysis because broad emission components in this region could be dominated by beam smearing and/or stacking artefacts (see Sections \ref{subsec:sinfoni_fitting} and \ref{sec:beam_smearing}).

There is a clear positive correlation between \sfrsd\ and $\sigma_b$, which can be well described by a power law (\mbox{$\sigma_b$ = $c_1~\Sigma_{SFR}^{c_2}$}). We fit only the 5 bin stacks and use orthogonal distance regression to account for both the spread in \sfrsd\ values within each bin and the errors on the $\sigma_b$ measurements. The best fit power law is shown by the purple line and filled error region in Figure \ref{fig:sfrsd_sig}, and is given by
\begin{equation}
\sigma_b~=~({241\pm 23}~\rm{km~s}^{-1})~\left(\frac{\Sigma_{SFR}}{\rm{M}_{\odot}~\rm{yr}^{-1}~\rm{kpc}^{-2}}\right)^{{0.30\pm 0.11}}
\label{eqn:sfrsd_sig_eqn}
\end{equation}

The errors on the best fit parameters and the error region around the best fit curve were calculated using bootstrapping. We randomly perturbed the location of each stack in the \sfrsd-$\sigma_b$ plane according a 2D Gaussian distribution with dispersion in the two dimensions given by the errors on \sfrsd\ and $\sigma_b$, and then re-fit the \mbox{\sfrsd-$\sigma_b$} relation using the perturbed values. This process was repeated 100 times. The quoted errors on the normalisation and power law index represent the 1$\sigma$ ranges in the parameters obtained from the 100 bootstraps, and the purple shaded region in Figure \ref{fig:sfrsd_sig} indicates the 16th-84th percentile range of the 100 best fit curves at each \sfrsd\ value.

We confirm that the \sfrsd-$\sigma_b$ correlation is not an artefact of beam smearing (which artificially increases the $\sigma_b$ in the central regions of galaxies where \sfrsd\ is often also the highest) by constructing and fitting another set of stacks, excluding nuclear spaxels (with galactocentric radius less than 3 kpc). Because the number of available spaxels is reduced, we also reduce the number of bins to four. The outflow component is not detected in the lowest \sfrsd\ stack, but the $\sigma_b$ values measured for the remaining 3 stacks are consistent (within the 1$\sigma$ errors) with the \sfrsd-$\sigma_b$ relation measured from the full set of spaxels.

It is important to consider the fact that separating the star formation and outflow components becomes more difficult as the $\sigma_b$ and/or BFR decrease. We employ a forward modelling technique (described in Appendix \ref{sec:forward_modelling}) to investigate how robustly we can recover the intrinsic parameters of the outflow component in different regions of parameter space. In summary, we find that the true BFR for the below median stack could be a factor of $\sim$2 higher than measured, but is at least a factor of 2 lower than the BFRs of the high \sfrsd\ stacks. Our modelling also indicates that the true $\sigma_b$ for the below median \sfrsd\ stack may be as high as $\sim$180~\kms, which is comparable to the $\sigma_b$ values measured for the three stacks at \mbox{0.25 -- 0.45~\myrkpc}, and suggests that the \mbox{$\sigma_b$-\sfrsd} relation may flatten at low \sfrsd.

\begin{figure*}
\centering
\includegraphics[scale=0.7, clip = True, trim = 10 10 10 10]{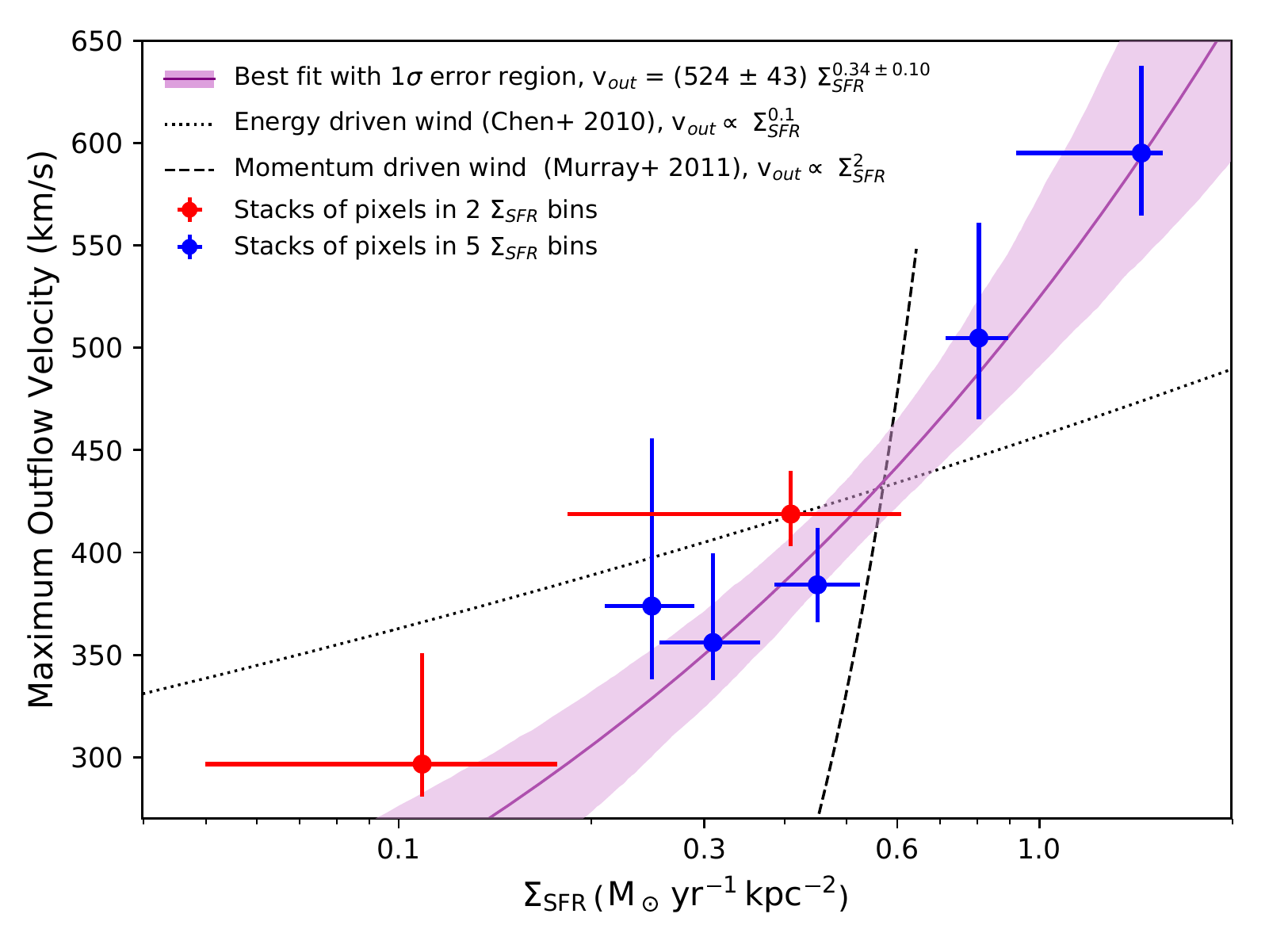}
\caption{Estimated outflow velocity \vout\ as a function of \sfrsd. The dotted and dashed lines show the predicted scalings between \sfrsd\ and \vout\ for energy driven and momentum driven outflow models, respectively. The scaling for our data lies between the two predictions, suggesting that the outflows may be driven by a combination of mechanical energy and momentum transport. \label{fig:sfrsd_vout}}
\end{figure*}

\subsection{Outflow Velocity}\label{subsec:vout}
We use the measured outflow component kinematics to investigate how the outflow velocity varies with \sfrsd. \citet{Heckman00} reported that if gas is deposited into an outflow at approximately zero velocity and is then accelerated outwards, the maximum (terminal) velocity of the outflow is roughly \mbox{$\Delta v - \textrm{FWHM}_{broad}$/2} \citep[see also][]{Rupke05, Veilleux05}. \citet{Genzel11} adopted a slightly different definition for the outflow velocity \mbox{(\vout~$\sim \Delta v - 2 \sigma_b$)}, which brings the outflow velocities they measure from emission line spectra into better agreement with outflow velocities measured from absorption line spectra, for galaxies with similar stellar masses and SFRs. We adopt the \citet{Genzel11} definition for \vout, but find that the slope of the \sfrsd-\vout\ relation is independent of the chosen prescription for \vout\ (see discussion below).


The velocity shift of the outflow component relative to the star formation component is always $<$~70~\kms\ (see e.g. Figure \ref{fig:sample_halves_fits}), and therefore the outflow velocity is almost proportional to $\sigma_b$. Consequently \vout, like $\sigma_b$, is positively correlated with \sfrsd\ (shown in Figure \ref{fig:sfrsd_vout}). We fit a power law relation to the \sfrsd\ and \vout\ values of the 5 bin stacks, and obtain 
\begin{equation}
v_{\rm out}~=~({524\pm 43}~\rm{km~s}^{-1})~\left(\frac{\Sigma_{SFR}}{\rm{M}_{\odot}~\rm{yr}^{-1}~\rm{kpc}^{-2}}\right)^{{0.34\pm 0.10}}
\label{eqn:sfrsd_vout_eqn}
\end{equation}

\begin{sloppypar}
The outflow velocities for the stacks at \mbox{\sfrsd~$>$~0.2~\myrkpc} range from \mbox{350-600~\kms}; in good agreement with the results of \citet{NMFS18b} who stacked global emission line spectra of galaxies with star formation driven outflows at 0.6~$<~z~<$~2.7 and found a typical outflow velocity of 450~\kms. Our outflow velocities are also similar to outflow velocities measured for galaxies with similar SFRs in various absorption line studies \citep[e.g.][]{Weiner09, Martin12, Kornei12, Heckman16}.
\end{sloppypar}

If we apply the \vout\ definition adopted by \citet{Heckman00}, \citet{Rupke05} and \citet{Veilleux05}, the normalisation in Equation \ref{eqn:sfrsd_vout_eqn} changes to 325~$\pm$~38~\kms\ -- a factor of 1.6 lower than derived using the \citet{Genzel11} definition. The best fit power law index is the same regardless of the adopted definition for \vout, and is in excellent agreement with many studies reporting similar power law scalings between \vout\ and either \sfrsd\ or SFR, across a range of redshifts \citep[e.g.][]{Martin05, Weiner09, Martin12, Kornei12, Heckman15, Heckman16}. 

In Figure \ref{fig:sfrsd_vout} we also compare our derived \vout-\sfrsd\ scaling to predictions from different star formation driven outflow models. The dotted line shows the best fit to our data for a scaling of \mbox{\vout~$\propto$ \sfrsd$^{0.1}$}, which is predicted for outflows driven by energy from supernova explosions \citep[e.g.][]{Strickland04, Chen10}, and the dashed line shows the best fit to our data for a scaling of \mbox{\vout~$\propto$ \sfrsd$^2$}, which is predicted for outflows driven by momentum transport through radiation pressure \citep[e.g.][]{Murray11, Kornei12}. The observed relationship between \vout\ and \sfrsd\ for our stacks lies between the two model predictions, suggesting that the outflows in the SINS/zC-SINF AO star forming galaxies may be driven by a combination of energy from supernova explosions and radiation pressure from massive stars. 

\subsection{Mass Loading Factor $\eta$}\label{subsec:mass_loading}
The mass loading factor $\eta$ is defined as the ratio of the mass outflow rate to the star formation rate, and is an important parameter governing the strength of feedback in cosmological simulations (see further discussion in Section \ref{subsec:mass_budget}). $\eta$ can be estimated from the outflow velocity and the \Ha\ broad-to-narrow ratio, following the method described in \citet{Genzel11} and \citet{Newman12_406690}. The model assumes that the outflow velocity and mass outflow rate are constant. The narrow component of the line emission is assumed to be associated with photoionized \HII\ region gas, while the broad component is assumed to be associated with photoionized gas in the outflow. Under these conditions (following \citealt{Newman12_global}), the mass outflow rate $\dot{M}_{\rm out}$ can be derived as follows:
\begin{align}
\dot{M}_{\rm out}~(\textrm{g~s}^{-1}) &= \frac{1.36 m_H}{\gamma_{\rm H\alpha} n_e} \left(\frac{v_{\rm out}}{R_{\rm out}} \right) L_{\rm H\alpha, broad} \label{eqn:mdot_out}
\end{align}
\noindent where 1.36$m_H$ is the effective nucleon mass for a 10 per cent helium fraction, $\gamma_{\rm H\alpha}$~=~3.56~$\times$~10$^{-25}$~erg~cm$^3$~s$^{-1}$ is the \Ha\ emissitivity at \mbox{T = 10$^4$K}, $n_e$ is the local electron density in the outflow, \rout\ is the maximum (deprojected) radial extent of the outflow, and $L_{\rm H\alpha, broad}$ is the \Ha\ luminosity of the outflow component. The SFR is given by
\begin{align}
\textrm{SFR~(g~s}^{-1}) &= \frac{1.99\times 10^{33}}{3.15 \times 10^7}\left(\frac{L_{\rm H\alpha, narrow}}{2.1 \times 10^{41}} \right) \label{eqn:sfr_eqn}
\end{align}
\noindent where $L_{\rm H\alpha, narrow}$ is the \Ha\ luminosity of the narrow component. The first term of Equation \ref{eqn:sfr_eqn} is the number of grams per solar mass divided by the number of seconds per year, and converts the SFR from units of $M_\odot$~yr$^{-1}$ to g~s$^{-1}$. Equations \ref{eqn:mdot_out} and \ref{eqn:sfr_eqn} can be combined to calculate the mass loading factor as follows:
\begin{align}
\eta &= \left( \frac{3.15 \times 2.1 \times 10^{48}}{1.99 \times 10^{33}} \right) \frac{1.36 m_H}{\gamma_{\rm H\alpha} n_e} \left( \frac{H\alpha_{\rm broad}}{H\alpha_{\rm narrow}} \right) \frac{v_{\rm out}}{R_{\rm out}}
\end{align}

In the following sub-sections we discuss the assumption that the outflowing gas is photoionized (Section \ref{subsubsec:ionization}), motivate our choice of electron density and outflow extent (Sections \ref{subsubsec:electron_density} and \ref{subsubsec:outflow_extent}), and analyse the variation in $\eta$ as a function of \sfrsd\ (Section \ref{subsubsec:eta_sfrsd}).

\subsubsection{Ionization Mechanisms}\label{subsubsec:ionization}
The assumption that both components are primarily photoionized is justified by the measured emission line ratios (\mbox{\NIIHa\ = 0.12$\pm$0.02} for the narrow component and \mbox{\NIIHa\ = 0.27$\pm$0.04} for the broad component of the above median \sfrsd\ stack), which are consistent with stellar photoionization \citep{Baldwin81, Ke01a}. The low \NIIHa\ ratio of the outflow component is surprising, because star-formation driven outflows in the local universe often show signatures of strong shocks \citep[e.g.][]{Sharp10, Rich11, Soto12b, Ho14}. In our z$\sim$2 star forming galaxies, the \NIIHa\ ratio of the outflow component is enhanced by 0.35 dex compared to the narrow component, which can be explained by a 30-40\% contribution from shock excitation \citep[see e.g.][]{Rich11, Yuan12}.

\citet{Sharp10} argue that the shock-like line ratios in local star formation driven outflows are the result of bursty star formation. By the time the energy from supernova explosions is able to dislodge gas from the disk and launch winds, most of the massive stars have died. It is possible that our $z\sim$~2 galaxies are experiencing extended periods of star formation, so that stellar photoionization dominates over shock excitation even after winds have been launched. However, a more likely explanation for the low shock fraction is that we are simply probing material close to the galaxy disks where the large scale stellar radiation field is strong. Low shock fractions are observed close to the disks of several prototypical starburst driven superwind galaxies including M82, NGC~1482 and NGC~253 \citep{Shopbell98, Veilleux02, Westmoquette11}.

\subsubsection{Electron Density}\label{subsubsec:electron_density}
The electron density determines the constant of proportionality between the broad \Ha\ luminosity and the outflow mass in Equation \ref{eqn:mdot_out}, but is notoriously difficult to measure. In principle, the electron density in the outflow can be calculated from the ratio of the broad component amplitudes of the \SII\ doublet lines (\SII$\lambda$6716/\SII$\lambda$6731; \citealt{Osterbrock06}). However, the \SII\ line emission in our stacks is not strong enough to independently constrain the amplitudes of the narrow and broad components. Previous studies have faced similar issues, leading to large uncertainties on the electron densities and the resulting mass loading factors. For example, \citet{Newman12_global} measured an electron density of \mbox{10$^{+590}_{-10}$ cm$^{-3}$} in the broad component for a stack of 14 galaxies with star formation driven outflows from the SINS/zC-SINF survey. Other recent studies of outflows at high redshift by \citet{Freeman17} and \citet{Leung17} were unable to constrain the electron density and adopted the \citet{Newman12_global} value.

Important progress was made by \citet{NMFS18b}, who performed the most accurate measurement to date of the electron density in star formation driven ionized gas outflows at high redshift. They started with a sample of 599 galaxies from the \mbox{SINS/zC-SINF} and \ktd\ surveys, which includes many of our galaxies and covers the main sequence of star forming galaxies at \mbox{0.6 $<~z~<$ 2.7} over a stellar mass range of \mbox{9.0 $<$ log(M$_*$/M$_\odot$) $<$ 11.7}. They stacked the spectra of 33 galaxies with individual high S/N detections of star formation driven outflows, and fit two components to each of the \SII\ lines in the stacked spectrum. They measured electron densities of \mbox{76$^{+24}_{-23}$ cm$^{-3}$} and \mbox{380$^{+249}_{-167}$ cm$^{-3}$} for the narrow and outflow components, respectively. The outflow component is denser than the star formation component, providing further evidence to suggest that the outflowing material may be shocked. The \citet{NMFS18b} results are consistent with several other studies of ionized gas outflows that have also found the outflowing gas to be denser than the \HII\ region gas \citep[e.g.][]{Arribas14, Ho14, Perna17, Kakkad18}. We therefore adopt an outflow electron density of 380~cm$^{-3}$. There are no observational constraints on how the electron density in the outflow varies as a function of \sfrsd, so we assume that it is constant.

\begin{figure*}
\centering
\includegraphics[scale=0.7, clip = True, trim = 10 15 10 10]{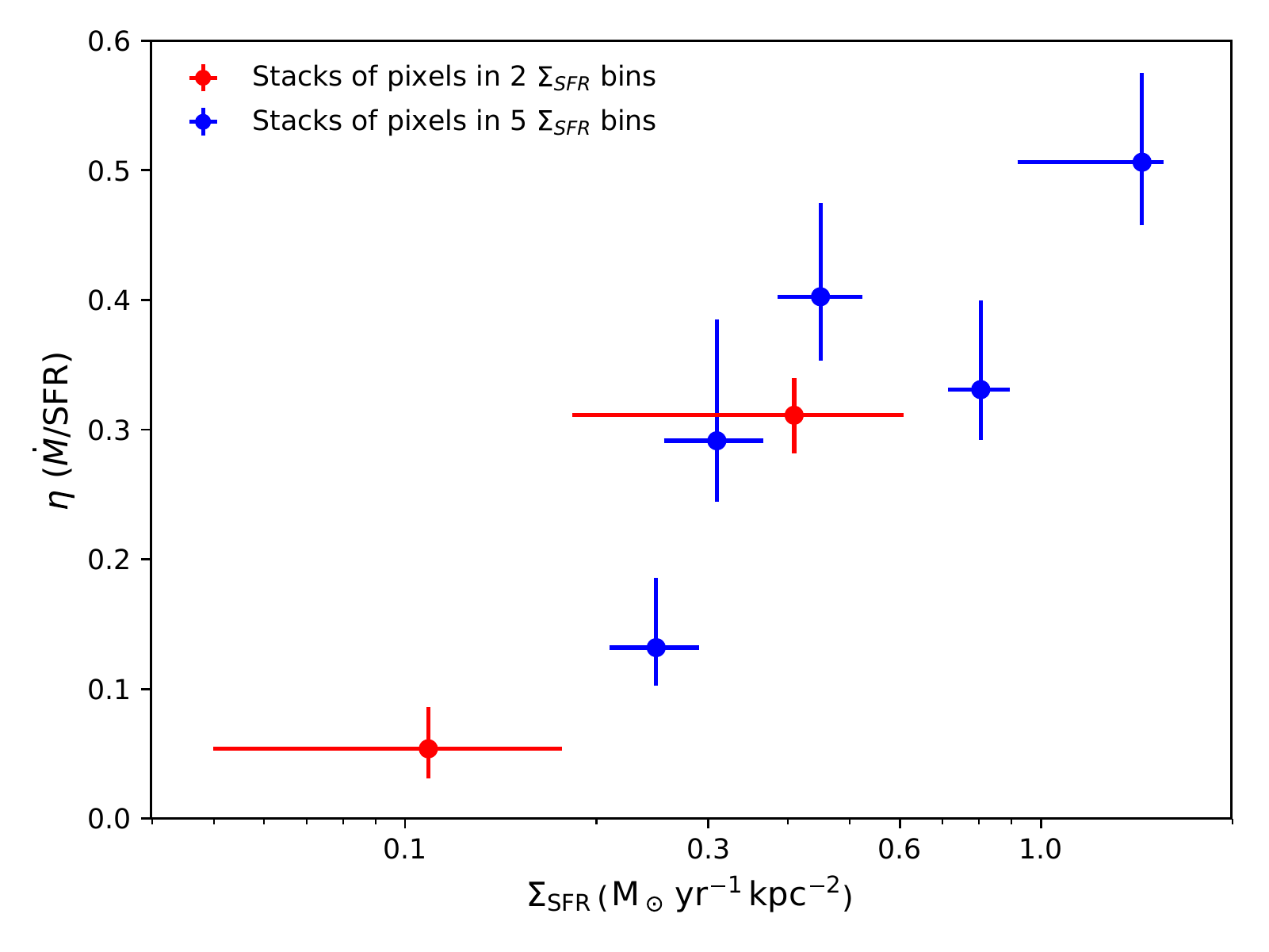}
\caption{Estimated mass loading factor $\eta$ as a function of \sfrsd. $\eta$ is inversely proportional to the electron density in the outflow component, which we assume here to be 380~cm$^{-3}$. The derived $\eta$ values are significantly lower than what is typically assumed for stellar feedback in cosmological simulations, suggesting that a significant fraction of the outflowing mass must be in other gas phases. \label{fig:sfrsd_eta}}
\end{figure*}

\subsubsection{Radial Extent of the Outflow}\label{subsubsec:outflow_extent}
The \vout/\rout\ term in Equation \ref{eqn:mdot_out} is the inverse of the dynamical time of the outflow (see e.g. \citealt{Veilleux05}), and \rout\ is the radial extent of the outflow from the point of launch. For the stacking analysis presented in this paper, the correct choice of \rout\ is not clear, because the stacks combine many spectra which trace outflows at different distances from their point of launch. Therefore, we explore two extreme cases to obtain upper and lower limits on \rout, and adopt an intermediate value in our subsequent calculations.

To obtain a lower limit on \rout, we assume that the outflowing material is always observed close to where it was launched. We take the minimum \rout\ to be the typical half width at half maximum of the PSF, which for our sample is 0.7~kpc \citep[see also][]{Newman12_406690}.

To obtain an upper limit on \rout, we assume that the outflowing material could have been launched from anywhere within the galaxies. The maximum galactocentric radius to which broad emission is observed in the SINS/zC-SINF AO galaxies is 2.6 kpc \citep{Newman12_global}, so we take that as the upper limit on \rout.

For our subsequent analysis we adopt the average of the lower and upper limits: \rout~=~1.7~kpc. As for $n_e$, we assume that \rout\ is independent of \sfrsd.

\subsubsection{Trends with \sfrsd}\label{subsubsec:eta_sfrsd}
Figure \ref{fig:sfrsd_eta} shows the $\eta$ values calculated for our stacks. The mass loading factor scales linearly with both BFR and \vout\, and consequently shows a clear positive correlation with \sfrsd. The positive correlation between \sfrsd\ and $\eta$ is in agreement with results from other studies of neutral and ionized gas outflows \citep[e.g.][]{Chen10, Newman12_global, Arribas14}, but is in tension with models which predict that $\eta$ should be inversely correlated with \sfrsd\ \citep[e.g][]{Hopkins12, Creasey13, Lagos13, Li17}. 

The errors on the $\eta$ values shown in Figure \ref{fig:sfrsd_eta} do not include the $\sim$~50\% errors on \rout\ and $n_e$, which would translate to a $\sim$~70\% uncertainty on $\eta$. If \rout\ and $n_e$ are independent of \sfrsd\ (as we have assumed), then the \textit{shape} of the \mbox{\sfrsd-$\eta$} relationship is independent of the chosen \rout\ and $n_e$ values, but there is a $\sim$~70\% error on the normalisation of the relationship (the average $\eta$). However, if \rout\ and $n_e$ vary as a function of \sfrsd, then the true shape of the relationship between \sfrsd\ and $\eta$ may be different.

We note that even at \mbox{\sfrsd\ $>$ 0.3~\myrkpc}, the $\eta$ values are relatively low \mbox{($\eta\ \sim$~0.3-0.5)}, suggesting that the ionized gas outflow rates are considerably smaller than the SFRs of the clumps driving the outflows. \citet{NMFS18b} found similarly low galaxy-integrated mass loading factors for star formation driven outflows at \mbox{0.6 $<$ z $<$ 2.7} in the \ktd\ survey. \citet{Freeman17} found higher mass loading factors of \mbox{0.64-1.4} for star formation driven outflows in galaxies with similar stellar masses and redshifts to the galaxies in our sample (\mbox{9.8 $<$ log(M$_*$/M$_\odot$) $<$ 10.7} at \mbox{1.37 $<$ z $<$ 2.61}), but this difference is due to the fact that they assumed an electron density of 50~cm$^{-3}$. If we re-calculate the mass loading factors for their sample assuming $n_e$~=~380~cm$^{-3}$, the mass loading factors decrease to $\eta$~=~0.08-0.18. Likewise, \citet{Newman12_global} found $\eta \sim$~2 for a very similar sample of galaxies to the one used in this paper, assuming $n_e$~=~50~cm$^{-2}$. If we re-scale this $\eta$ value for $n_e$~=~380~cm$^{-3}$, we obtain $\eta \sim$~0.26. 

\begin{figure*}
\centering
\includegraphics[scale=0.59, clip = True, trim = 10 10 0 0]{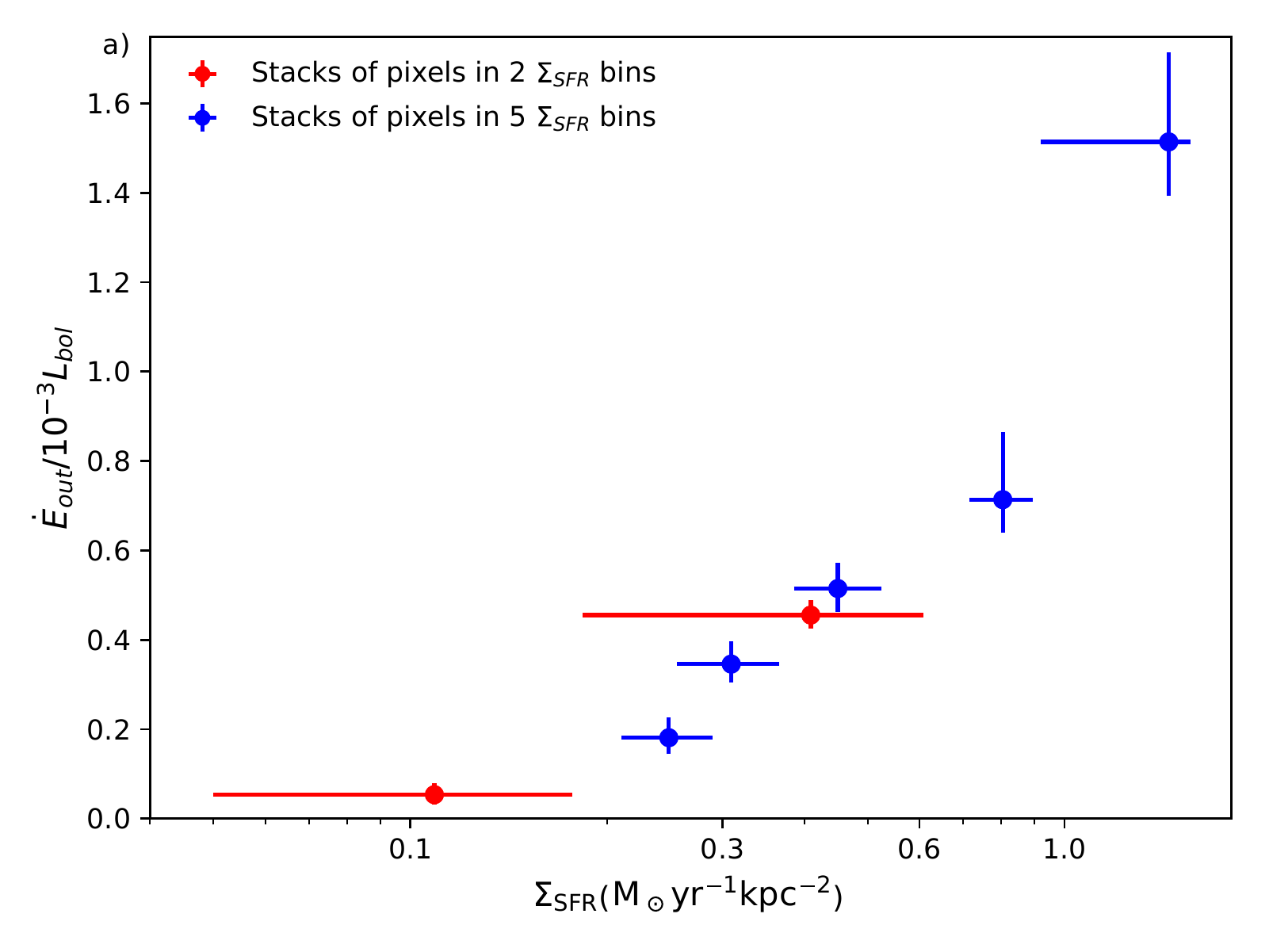}\includegraphics[scale=0.59, clip = True, trim = 10 10 10 0]{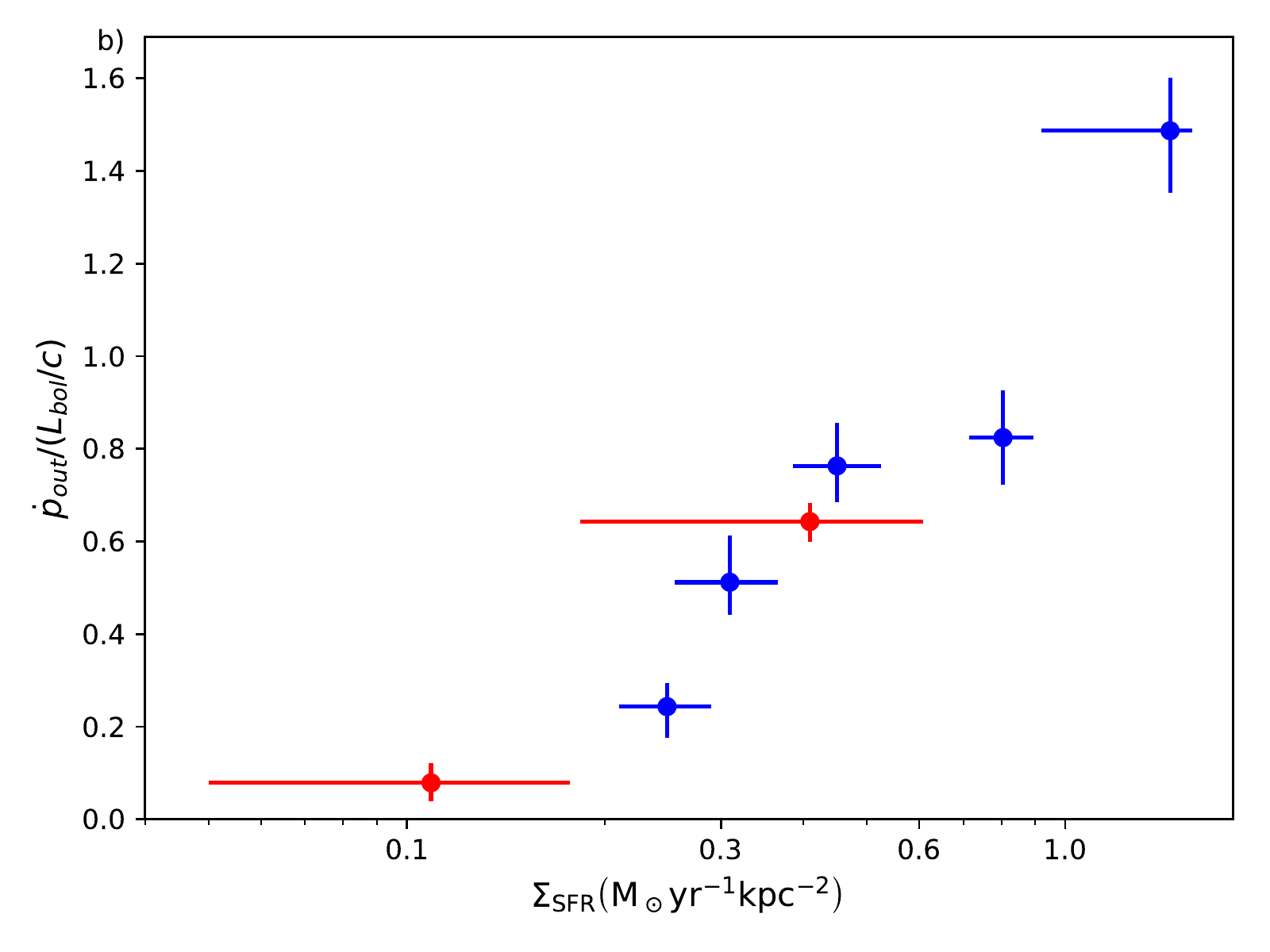}
\caption{Panel a): Measured energy outflow rate $\dot{E}_{\rm out}$ as a fraction of the energy outflow rate predicted by \citet{Murray05} for energy driven winds (10$^{-3}$L$_{\rm bol}$). Panel b): Measured momentum outflow rate $\dot{p}_{\rm out}$ as a fraction of the predicted momentum outflow rate for momentum driven winds (L$_{\rm bol}$/c). \label{fig:sinfoni_outflow_energetics}}
\end{figure*}

We emphasise that the normalisation of the mass loading factor is impacted by uncertainties on the electron density and the radial extent of the outflow. However, if the ionized gas mass loading factors are indeed so low, the outflows would not appear to be able to remove enough gas to explain the low baryon fractions of low mass halos \citep[e.g.][]{Baldry08, Moster13, Moustakas13, Behroozi13}. We have measured only the warm ionized gas phase of the outflows, and the overall mass loading factors could be higher if most of the outflowing mass is in other gas phases (see also discussion in \citealt{NMFS18b}). This will be discussed further in Section \ref{subsec:mass_budget}.

\subsection{Outflow Energetics}
In the absence of an AGN, galaxy scale outflows are generally thought to be driven by either the energy released through supernova explosions and radiation from young stars, or by the transport of momentum by means of radiation pressure, supernovae and stellar winds. Assuming that 1 per cent of the bolometric luminosity of the stars is released as energy in supernova explosions, and that 10 per cent of this energy is able to couple to the ISM, \citet{Murray05} predict that for an energy driven wind, the energy outflow rate should be $\dot{E}_{\rm out} \sim$~10$^{-3}$~L$_{\rm bol}$, whereas for a momentum driven wind, the momentum outflow rate should be $\dot{p}_{\rm out} \sim$~L$_{\rm bol}$/c. We assume that \mbox{$L_{\rm bol}$ $\sim$ SFR~$\times$~10$^{10}$~L$_\odot$} \citep{Kennicutt98}.

The energy and momentum outflow rates for each of the stacks can be calculated as follows:
\begin{eqnarray}
\dot{E}_{\rm out} &=& \frac{1}{2} \dot{M} v_{\rm out}^2  = \frac{1}{2} \eta \times \textrm{SFR} \times v_{\rm out}^2  \\
\dot{p}_{\rm out} &=& \dot{M} v_{\rm out} = \eta \times\ \mathrm{SFR} \times v_{\rm out}
\end{eqnarray}
Figure \ref{fig:sinfoni_outflow_energetics} shows the ratios of the calculated energy and momentum outflow rates for each of the stacks to the predicted energy and momentum outflow rates from \citet{Murray05}. For the stacks with \mbox{\sfrsd\ $<$ 1.0~\myrkpc}, the ratios are in the following ranges:
\begin{eqnarray}
\frac{\dot{E}_{\rm out}}{10^{-3}{\rm L}_{\rm bol}} &=& 0.05-0.71~\times~\frac{380~{\rm cm}^{-3}}{{\rm n}_e} \frac{1.7~{\rm kpc}}{{\rm R}_{\rm out}} \\
\frac{\dot{p}_{\rm out}}{{\rm L}_{\rm bol}/{\rm c}} &=& 0.09-0.85~\times~\frac{380~{\rm cm}^{-3}}{{\rm n}_e} \frac{1.7~{\rm kpc}}{{\rm R}_{\rm out}} 
\end{eqnarray}
The ratios do not exceed 1, suggesting that the current star formation activity is sufficient to power the observed outflows and that no additional energy source is required. However, the ratios show a strong correlation with \sfrsd\ which is not explained by the \citet{Murray05} models. 

For the highest \sfrsd\ stack, the energy and momentum outflow rates exceed the predicted rates by a factor of 1.5. There are several possible explanations for this discrepancy. A significant amount of the energy or momentum driving the outflows may come from source(s) which are not accounted for in the \citet{Murray05} model, such as cosmic rays \citep[see e.g.][]{Ruszkowski17, Girichidis18}. The SFR measured from the narrow component of the \Ha\ emission may not be representative of the SFR at the time and point of launch of the outflows, either because the SFR has changed over time, or because the outflows have propagated away from their launch points (which is the case for one of the outflows in ZC406690; \citealt{Newman12_406690}). Alternatively, one or more of the assumptions made in our calculations may be incorrect: the adopted electron density or extent of the outflow may be too low, the majority of the outflowing gas may be collisionally excited rather than photoionized (which would change the coefficients in the outflow mass calculation and reduce $\eta$ by a factor of $\sim$2; \citealt{Genzel11}), or the emission lines may be partially broadened by mechanisms other than outflows (such as shocks or turbulent mixing layers; see further discussion in Section \ref{subsec:shocks_mixing}).

\section{Discussion}\label{sec:discussion}
\subsection{Escape Fraction and Mass Budget}\label{subsec:mass_budget}
In order to gauge the impact of the star formation driven outflows on the stellar mass growth and structural evolution of their host galaxies, we investigate whether any of the outflowing material is travelling fast enough to escape the galaxy halos. The halo escape velocity is approximately three times the galaxy circular velocity \citep[e.g.][]{Weiner09}. We calculate the `characteristic' circular velocity for each \sfrsd\ bin by assigning each spaxel the circular velocity of its host galaxy (given in \citealt{NMFS18a}), and then averaging the circular velocities assigned to all the spaxels in the relevant bin. The characteristic circular velocities (halo escape velocities) decrease from $\sim$230~\kms\ (690~\kms) at the lowest \sfrsd\ to $\sim$180~\kms\ (540~\kms) at the highest \sfrsd. The highest \sfrsd\ bin is the only one for which the outflow velocity (595~\kms) exceeds the characteristic halo escape velocity. Therefore, if the outflows are spherically symmetric, very little of the outflowing material is likely to escape from the galaxy halos.

If the outflows are biconical and perpendicular to the galaxy disks \citep[as suggested by observations of higher outflow velocities in face on galaxies than edge on galaxies; e.g.][]{Heckman00, Chen10, Kornei12, Newman12_global, Bordoloi14}, the outflow velocities will be under-estimated by a factor of 1/cos(\textit{i}), which is equivalent to the inverse of the typical galaxy axis ratio. We calculate characteristic axis ratios for each of our stacks using the same method applied to calculate the characteristic circular velocities. We find that if the outflows are biconical, the outflowing material in the two highest \sfrsd\ bins would have sufficient velocity to escape the galaxy halos. 

We note that if the outflowing material is still being accelerated, the terminal velocities will be higher than the measured outflow velocities and more material will be able to escape from the halos. On the other hand, if the outflows are ballistic, the outflow velocity will decrease over time and less material will be able to escape. Better observational constraints on the geometry and velocity structure of star formation driven outflows are required to more accurately determine the fate of the outflowing material.

We conclude that the majority of the ionized gas in the outflows is likely to decelerate and be re-accreted onto the galaxy disks. The re-accretion of gas launched in outflows is important for the chemical evolution of galaxies because outflows carry significant amounts of heavy metals \citep[e.g.][]{Larson74, Tremonti04, Finlator08, Peeples14, Zahid14}. Galaxy formation simulations with star formation feedback suggest that the timescale for re-accretion of the outflowing material is approximately 1-2 Gyr \citep[e.g.][]{Oppenheimer10, Brook14}. 

The mass loading factors derived in this paper are quite low; \mbox{$\eta$ $\sim$ (0.3 - 0.5)~$\times$~380~cm$^{-3}$/$n_e$}. $\eta$ is proportional to \vout, so if the outflows are biconical the maximum $\eta$ could increase to $\sim$~0.7, but this would only partially resolve the tension with cosmological simulations which typically require $\eta \ga$~1 to reproduce the low baryon fractions of low mass halos \citep[e.g.][]{Finlator08, Dave11, Hopkins12, Vogelsberger14, Muratov15}. 

The discrepancy between the predicted and measured mass loading factors can potentially be resolved by considering the mass contained in other phases of the outflows. Multi-wavelength observations of star formation and AGN driven outflows, both locally and at high redshift, suggest that the ionized gas phase accounts for only a small fraction of the total mass and energy expelled in outflows. The ionized gas outflow in M82 contains only $\sim$1-2\% of the mass carried in the neutral and molecular phases of the outflow \citep{Shopbell98, Walter02, Contursi13, Leroy15}. Similary, in the local starburst/quasar ultra-luminous infrared galaxy Mrk~231, the ionized gas outflow rate is $\sim$4000 times lower than the molecular and neutral gas outflow rates \citep[e.g.][]{Cicone14}. In general, AGN driven outflows appear to carry more mass and energy in the neutral and molecular gas phases than in the ionized gas phase \citep[e.g.][]{Rupke13, Carniani17, Fluetsch18, HerreraCamus19}. There is some evidence that in local star forming galaxies the ionized gas outflow rate may be comparable to the neutral and molecular gas outflow rates \citep[e.g.][]{Fluetsch18}, but larger samples of galaxies with observations in multiple outflow tracers are required to overcome the large uncertainties on the mass outflow rates. 

It is also important to consider the contribution of the hot X-ray emitting gas. This phase is very difficult to observe, but multi-phase simulations of star formation driven outflows predict that at distances of $\gtrsim$~1~kpc from the disk, the majority of the outflowing mass and energy will be carried by the hot phase \citep[e.g.][]{Li17, Fielding18, Kim18}. 

It is clear that the total mass outflow rates and gas recycling rates are strongly dependent on the multi-phase mass budget, geometry and velocity structure of the outflows (see also discussion in \citealt{NMFS18b}). Better constraints on these properties are therefore key to improving our understanding of the impact of star formation feedback on the growth and evolution of $z\sim$~2 star forming galaxies.

\subsection{Alternative Sources of Broad Emission}\label{subsec:shocks_mixing}
Throughout this paper we have assumed that the broad component of the \Ha\ line emission traces the bulk motion of ionized gas entrained in star formation driven outflows, and therefore that the velocity dispersion of the broad component is a direct tracer of the outflow velocity. In Section \ref{sec:broad_emission} we showed that the \Ha\ line width is strongly correlated with \sfrsd, indicating that the broad emission is related to star formation. On the other hand, the shape of the \Ha\ line is not correlated with either the $A_V$ or the galactocentric distance, indicating that scattered light and beam smearing are unlikely to be significant sources of broad emission in our galaxies (see also discussion in Appendix \ref{sec:beam_smearing}).

Absorption line studies indicate that outflows are ubiquitous in star forming galaxies at z$\sim$2 \citep[e.g.][]{Shapley03, Weiner09, Rubin10, Steidel10, Erb12, Kornei12, Bordoloi14}. It is therefore not surprising that we observe broad \Ha\ emission associated with the ionized phase of these outflows. However, it is important to consider the possibility that the kinematics of the outflowing gas may include significant contributions from turbulent motions as well as bulk flows.

The outflowing gas may collide with material in the ISM of the galaxies and trigger shocks. The broad component of the above median \sfrsd\ stack has an \NIIHa\ ratio of 0.27$\pm$0.04, which is significantly lower than would be expected for purely shock-excited gas \citep[e.g.][]{Allen08, Rich11}, but significantly higher than the 0.12$\pm$0.02 measured for the narrow component. This suggests that up to 30-40\% of the \Ha\ emission could be shock excited, with the remaining gas photoionized by young stars in the galaxy disks (see discussion in Section \ref{subsubsec:ionization}). \citet{NMFS18b} found that the material emitting the broad \Ha\ is denser than the \HII\ region gas, suggesting that the broad component may trace compressed clumps of ionized gas entrained within the wind fluid. If the outflowing gas is shock excited, the width of the broad component would reflect the shock velocity, but the shock velocity is expected to be similar to the outflow velocity \citep[see e.g.][]{Soto12b, Ho14, McElroy15}.

Broad emission could also arise from turbulent mixing layers at the interface between cold gas in the disk and hot wind fluid \citep[e.g.][]{Slavin93, Esquivel06, Westmoquette07, Westmoquette09, Westmoquette11, Wood15}. In this scenario, the width of the broad component would reflect the turbulent velocity rather than the outflow velocity. We cannot rule out the possibility that the outflow components in our stacks are broadened by turbulent mixing layers in the ISM. However, the \sfrsd-\vout\ scaling we measure in Section \ref{subsec:vout} is consistent with results from absorption line studies of star-formation driven outflows at low and high redshift \cite[e.g.][]{Martin05, Weiner09, Heckman16, Sugahara17}, which suggests that the kinematics of the outflowing gas are likely to be dominated by bulk flows, with only a minor contribution from turbulent motions.

\section{Summary and Conclusions}
\label{sec:conc}

We investigated the relationship between star formation activity and the incidence and properties of outflows on scales of 1-2 kpc in a sample of 28 star forming galaxies at $z\sim$~2~--~2.6 from the SINS/zC-SINF AO Survey. This work builds on previous studies of the relationship between \textit{global} galaxy properties and outflow properties in the SINS/zC-SINF AO sample \citep{Newman12_global}, and the relationship between the \textit{resolved} star formation and outflow properties of star forming clumps in 5 SINS/zC-SINF AO galaxies \citep{Genzel11, Newman12_406690}. With the aid of stacking we are able to probe not only the actively star forming clump regions which have been studied previously, but also the fainter inter-clump regions, spanning a factor of $\sim$50 in \sfrsd. 

We divided the spaxels from the 28 datacubes into bins of different physical properties (SFR, \sfrsd, stellar mass surface density \smsd, \sfrsd/\smsd, $A_V$ and galactocentric distance), and stacked the spectra of the spaxels in each bin to obtain high signal to noise \Ha\ line profiles. The \Ha\ profiles were used to simultaneously probe the star formation (from the narrow component of the line) and the outflows (which, when present, produce an additional broader line emission component). The width of the outflow component is a tracer of the outflow velocity, and the flux of the outflow component is a tracer of the mass in the outflow. Our main results are as follows:

\begin{enumerate}
 \item The width of the \Ha\ line is most strongly dependent on the level of star formation (probed by the SFR and \sfrsd), supporting the notion that the observed broad emission is associated with star formation driven outflows. \smsd\ may also play a role in governing the incidence and properties of the outflows.
 \item The outflow component contains an average of $\sim$45\% of the \Ha\ flux emitted from the highest \sfrsd\ regions, but is less prominent at lower \sfrsd.
 \item The outflow velocity scales as \mbox{\vout~$\propto$~\sfrsd$^{0.34 \pm\ 0.10}$}. This scaling is shallower than the predicted \mbox{\sfrsd$^2$} dependence for outflows driven by momentum transport through radiation pressure, but steeper than the predicted \mbox{\sfrsd$^{0.1}$} dependence for outflows driven by kinetic energy from supernovae and stellar winds, suggesting that the observed outflows may be driven by a combination of these mechanisms. 
 \item The outflow velocity is lower than the halo escape velocity in all but the highest \sfrsd\ regions, indicating that the majority of the outflowing material will not be expelled but will decelerate and fall back onto the galaxy disks. Simulations suggest that this material will likely be re-accreted after 1-2 Gyr, contributing to the chemical enrichment of the galaxies.
 \item The mass loading factor $\eta$ increases with \sfrsd. The normalisation of $\eta$ is uncertain due to the large uncertainties on the radial extent and electron density of the outflowing material, but we find \mbox{$\eta~\sim$ 0.4 $\times$ (380~cm$^{-3}$/n$_e$) $\times$} \mbox{(1.7~kpc/R$_{\rm out}$)}. This may be in tension with cosmological simulations (which typically require $\eta~\ga$~1 to explain the low efficiency of formation of low mass galaxies), unless a significant fraction of the outflowing mass is in other gas phases and is able to escape the galaxy halos.
 \item In 6/7 stacks the current star formation activity is powerful enough to drive the observed outflows. The energy and momentum outflow rates for the highest \sfrsd\ stack exceed the predicted rates for star formation driven outflows by a factor of 1.5. This may indicate that other energy sources (such as cosmic rays) contribute significantly to driving the outflows, that the SFR has changed since the outflows were launched, that the outflows have propagated away from their point of launch, that the adopted electron density or extent of the outflow is too low, and/or that the emission lines are partially broadened by mechanisms other than outflows (such as shocks or turbulent mixing layers).
\end{enumerate} 

Our results confirm that \sfrsd\ is closely related to the incidence and properties of outflows on 1-2~kpc scales. In this paper we have only explored the average incidence and properties of outflows as a function of \sfrsd, which makes it difficult to draw strong conclusions on the relative importance of global and local galaxy properties in determining the properties of the outflows. In the future it will be important to investigate how much the outflow velocity and \Ha\ broad flux ratio vary at fixed local \sfrsd, and determine which local and/or global properties are responsible for driving these variations.

\acknowledgements
We thank the referee for their detailed feedback and useful suggestions which improved the clarity of this paper. RLD would like to thank Alice Shapley and Sylvain Veilleux for insightful discussions about this work. ST is supported by the Smithsonian Astrophysical Observatory through the CfA Fellowship. ESW acknowledges support by the Australian Research Council Centre of Excellence for All Sky Astrophysics in 3 Dimensions (ASTRO 3D), through project number CE170100013. This research made use of \textsc{Astropy}, a community-developed core Python package for Astronomy \citep{Astropy13, Astropy18}, \textsc{Matplotlib} \citep{Hunter07}, and \textsc{Statsmodels} \citep{Seabold10}.

\appendix

\section{Impact of Beam Smearing and Velocity Shifting Errors}\label{sec:beam_smearing}
We use dynamical models constructed with the IDL toolkit \textsc{DYSMAL} \citep{Davies11} to investigate whether a significant fraction of the broad emission in our stacks could originate from beam smearing and/or smearing introduced in the velocity shifting of the spectra. For each galaxy in our sample, we construct a rotating disk model which is tailored to the structural, kinematic and line emission properties of the galaxy (inclination, effective radius, sersic index, stellar mass, circular velocity, intrinsic velocity dispersion, total H$\alpha$ flux, and integrated \NIIHa\ ratio, derived as described in \citealt{NMFS18a}). Each rotating disk model is sampled onto a mock datacube with the same field of view, pixel scale and wavelength scale as our real SINFONI data, and smoothed spectrally and spatially to match the spectral line spread function and spatial PSF of the observations. Noise is added to each mock datacube based on the exposure time and the known shape of the SINFONI K band error spectrum.

For each mock datacube we create maps of the \Ha\ flux and kinematics, mask spaxels with \Ha~S/N~$<$~5, and shift the spectra of the remaining (unmasked) spaxels to zero velocity using the same methods applied to the real datacubes. We stack the unmasked spaxels from all the mock cubes using two different weighting schemes (1/rms weighting and no weighting), and also create stacks from just the spaxels within a 3 spaxel radius of the center of a galaxy. The four mock stacks are fit using the same MCMC fitting method applied to our data, except that we do not require $\sigma_b$~$>$~150~\kms. Instead, we only require that the outflow component is broader than the narrow component; i.e. \mbox{$\sigma_b$ - $\sigma_n$ $>$ 20~\kms}. There are no outflows in the mock cubes, and therefore any detected `outflow component' is purely the result of beam smearing and/or smearing introduced in the velocity shifting.

The maximum $\sigma_b$ measured for any of the four mock stacks is \mbox{$\sigma_b$ = 110~$\pm$~12~\kms}. Therefore, the 3$\sigma$ upper limit on the velocity dispersion of a broad component that could be primarily associated with beam smearing is 146~\kms. This is why we require the outflow components fit to our stacks to have $\sigma_b~>$~150~\kms.

\section{Using Forward Modelling to Investigate the Accuracy of the Recovered Outflow Parameters}\label{sec:forward_modelling}
\subsection{Summary of Methodology}
The analysis presented in this paper is based on the assumption that the star formation and outflow components of all the data stacks are intrinsically Gaussian, and that the parameters of these Gaussians can be robustly determined by performing two component Gaussian fitting. The fitting should indeed be robust when the wings of the outflow component are clearly detected above the wings of the narrow component and above the noise. The separation becomes more uncertain when the intrinsic $\sigma_b$ is low (making the outflow component difficult to distinguish from the star formation component), and/or the intrinsic BFR is low (bringing the flux per channel in the outflow component close to the noise level) \citep[see also][]{Genzel14, Freeman17}. We use forward modelling to understand the potential impact of this uncertainty on the \sfrsd-BFR and \sfrsd-$\sigma_b$ trends shown in Figures \ref{fig:sfrsd_fbroad} and \ref{fig:sfrsd_sig}. 

\begin{figure*}
\centering
\includegraphics[scale=0.9, clip = True, trim = 28 35 30 30]{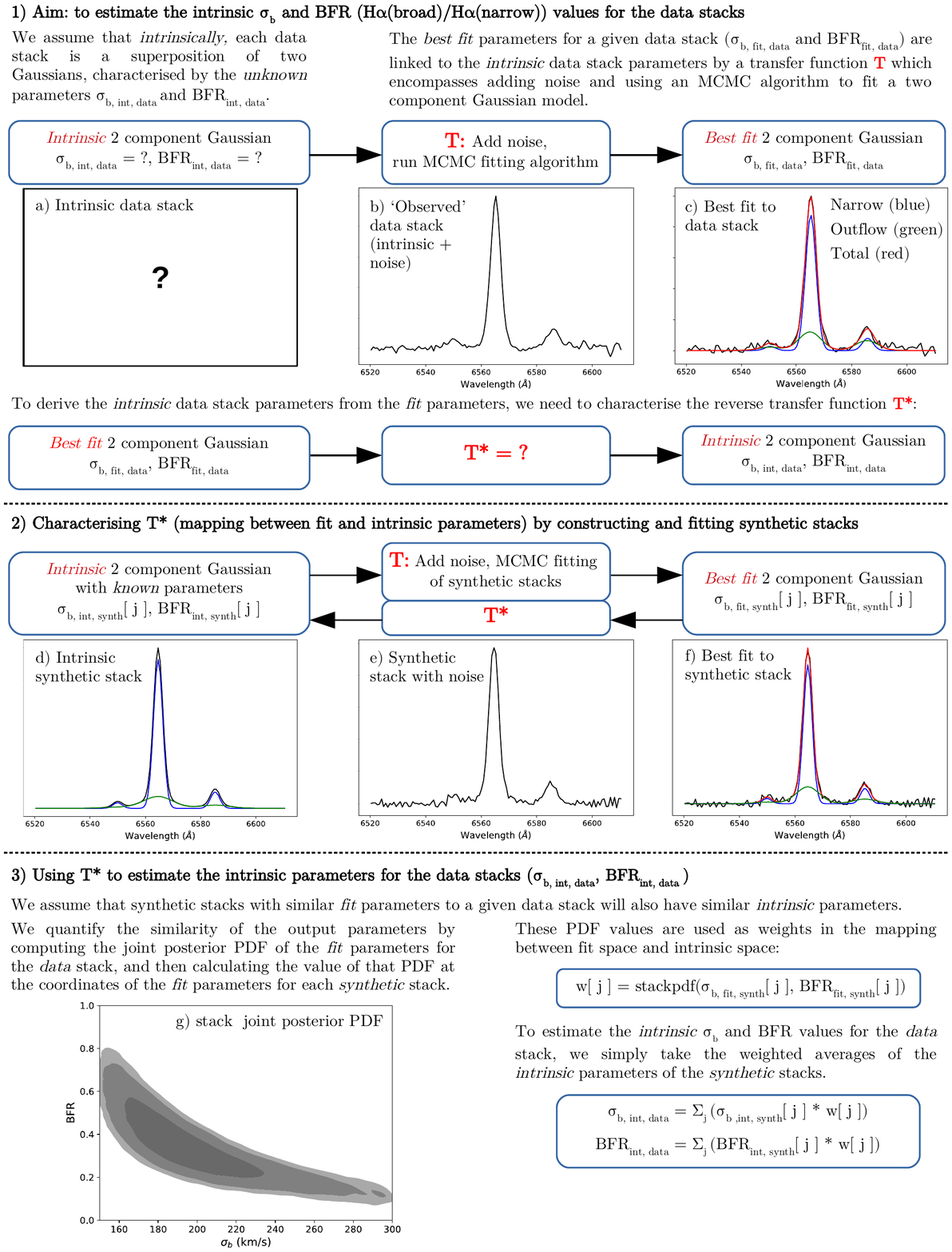}
\caption{Schematic describing the forward modelling technique used to estimate the intrinsic $\sigma_b$ and BFR of each data stack. \label{fig:forward_modelling}}
\end{figure*}

A schematic summarising the steps involved in the forward modelling is shown in Figure \ref{fig:forward_modelling}. In brief, we generate synthetic spectra with known BFR and $\sigma_b$, add noise to mimic the error spectra of the real data stacks, and fit the synthetic stacks with two Gaussian components using the same procedure applied to our data stacks. For each data stack, we identify synthetic spectra with similar fit parameters, and use the intrinsic parameters of these synthetic stacks to estimate the intrinsic parameters of the data stack. In performing this mapping, we do not simply take the synthetic spectrum with the closest fit parameters, or weight the synthetic spectra by the Euclidean distance between their fit parameters and the fit parameters of the data stack in BFR$_{\rm fit}$-$\sigma_{\rm b, fit}$ space, because these metrics do not account for the strong covariance between the BFR and $\sigma_b$ (seen in e.g. Figure \ref{fig:sample_halves_fits}). Instead, we use the joint BFR$_{\rm fit}$-$\sigma_{\rm b, fit}$ posterior probability distribution function (PDF) of the data stack to weight the synthetic spectra.

\subsection{Detailed Description}
The aim of the forward modelling (summarised in segment 1 of Figure \ref{fig:forward_modelling}) is to estimate the intrinsic $\sigma_b$ and BFR values for each of the data stacks. Throughout this paper, we assume that \textit{intrinsically} (if the noise was removed), each data stack is well explained by a superposition of two Gaussian components which are characterised by the \textit{unknown} parameters BFR$_{\rm int, data}$ and $\sigma_{\rm b, int, data}$. The actual data stacks analysed in this paper (such as the example shown in panel b) of Figure \ref{fig:forward_modelling}) are noisy realisations of the intrinsic stacks, where the noise is a combination of Gaussian noise from the instrument and uncertainties introduced during the stacking process. In our analysis, these noisy stacks were fit with two Gaussian components using the MCMC algorithm described in Section \ref{subsec:sinfoni_fitting}, yielding best fit parameters BFR$_{\rm fit, data}$ and $\sigma_{\rm b, fit, data}$ (panel c). 

A more mathematical expression of the relationship between the intrinsic and best fit parameters is as follows. Each intrinsic data stack, characterised by the unknown parameters BFR$_{\rm int, data}$ and $\sigma_{\rm b, int, data}$, is passed through a transfer function \textbf{T}, which maps between intrinsic space and fit space by adding noise, running an MCMC fitting algorithm, and returning the best fit parameters BFR$_{\rm fit, data}$ and $\sigma_{\rm b, fit, data}$. Therefore, to obtain the unknown intrinsic parameters BFR$_{\rm int, data}$ and $\sigma_{\rm b, int, data}$, we need to characterise the reverse transfer function \textbf{T*} which maps between fit space and intrinsic space. The relationship between the intrinsic and fit parameters and the forward and reverse transfer functions is shown by the two flowcharts in segment 1 of Figure \ref{fig:forward_modelling}.

To characterise \textbf{T*}, we construct a grid of 4200 synthetic stacks which mimic the data stacks but have \textit{known} intrinsic parameters (summarised in segment 2 of Figure \ref{fig:forward_modelling}). Like the data stacks, each synthetic stack is intrinsically a superposition of two Gaussian components. We construct synthetic stacks with intrinsic Gaussian parameters covering and extending beyond the range of best fit parameters measured for the data stacks: \mbox{0 $\leq$ BFR$_{\rm int, synth}$[j] $<$ 1.5} and \mbox{90 $\leq$ $\sigma_{\rm b, int, synth}$[j] (\kms) $<$ 300}. An example of an intrinsic synthetic stack and its constituent Gaussian components is shown in panel d) of Figure \ref{fig:forward_modelling}. 

We apply the forward transfer function \textbf{T} to each of the intrinsic synthetic stacks by adding noise to mimic the error spectra of the data stacks (panel e), and then fitting two Gaussian components using the same MCMC algorithm applied to the data stacks to obtain the best fit parameters \mbox{BFR$_{\rm fit, synth}$[j]} and \mbox{$\sigma_{\rm b, fit, synth}$[j]} (panel f). With both the intrinsic and best fit parameters for each of the synthetic stacks, it is possible to map between fit and intrinsic space, mimicking the reverse transfer function \textbf{T*}.

The final step is therefore to use the pairs of intrinsic and best fit parameters for the synthetic spectra to estimate the intrinsic parameters of each data stack (summarised in segment 3 in Figure \ref{fig:forward_modelling}). We assume that synthetic stacks with similar \textit{fit} parameters to a given data stack will also have similar \textit{intrinsic} parameters. We quantify the similarity in the output parameters between any given data stack and all the synthetic stacks by computing a non-parametric kernel density estimate of the joint posterior PDF between the fit parameters of the \textit{data} stack (BFR$_{\rm fit, data}$ and $\sigma_{\rm b, fit, data}$), and then calculating the value of that PDF at the coordinates of the fit parameters for each synthetic stack (\mbox{BFR$_{\rm fit, synth}$[j]}, \mbox{$\sigma_{\rm b, fit, synth}$[j]}). An example joint posterior PDF for one of the data stacks is shown in panel g). The calculated PDF values are used as weights (\textit{w}[j]) to apply the mapping from fit space to intrinsic space. 

To estimate the \textit{intrinsic} BFR of each data stack (BFR$_{\rm int, data}$), we simply take the weighted average of the \textit{intrinsic} BFRs of the synthetic spectra $\left(\Sigma_j \textrm{BFR}_{\rm int, synth}\textrm{[j]} \times w\textrm{[j]} \right)$. The same method is applied to estimate $\sigma_{\rm b, int, data}$. 

\subsection{Results}
\begin{figure*}
\centering
\includegraphics[scale=0.9, clip = True, trim = 0 10 0 10]{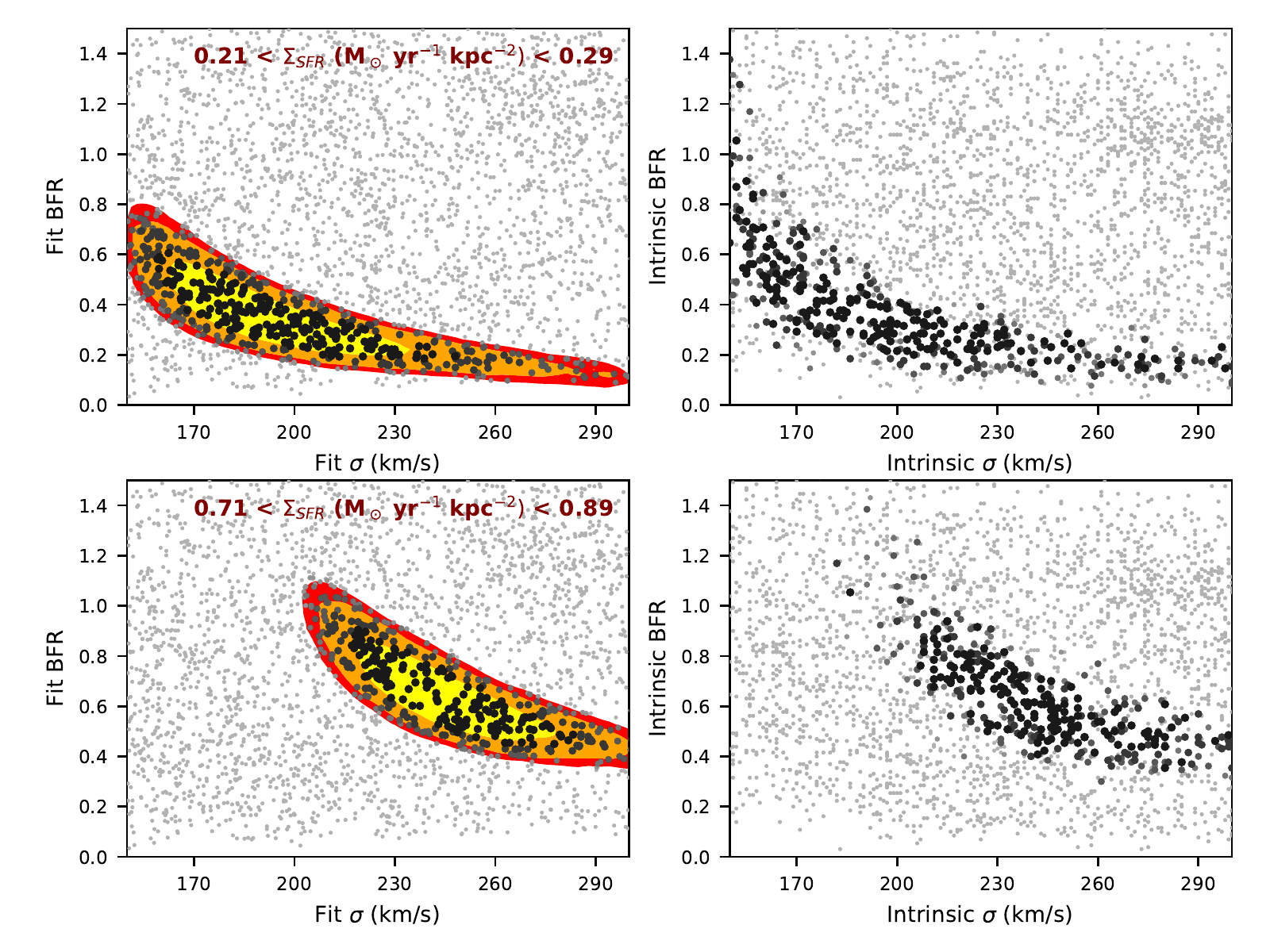}
\caption{Figure illustrating the mapping between fit and intrinsic space for a low \sfrsd\ stack (top row) and a high \sfrsd\ stack (bottom row). The yellow, orange and red shaded regions in the background of the left hand panels show the 1$\sigma$, 2$\sigma$ and 3$\sigma$ levels of the joint posterior PDF of the relevant data stack. The dots in the left hand panels show the \textit{best fit} $\sigma_b$ and BFR values for all the synthetic spectra, and the dots in the right hand panels show the \textit{intrinsic} $\sigma_b$ and BFR values for all the synthetic spectra. The two left hand panels have the same set of dots, as do the two right hand panels, but the size and color of the dots in each panel indicate the weights assigned to each synthetic spectrum in the mapping for the relevant data stack (large dots with dark colors have the highest weights). \label{fig:threshold_recovery_test}}
\end{figure*}

\begin{figure*}
\centering
\includegraphics[scale=0.59, clip = True, trim = 10 10 0 10]{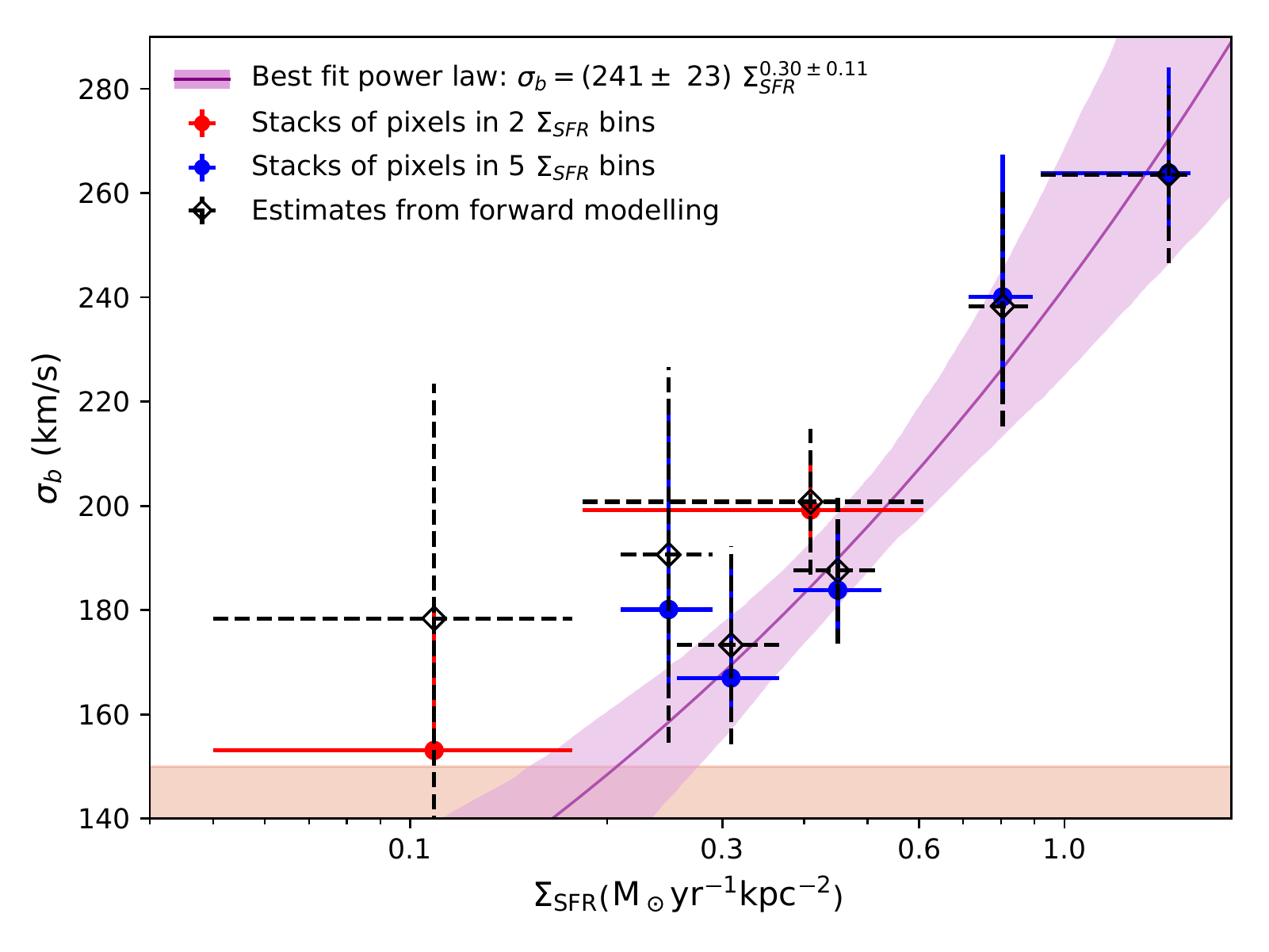}\includegraphics[scale=0.59, clip = True, trim = 10 10 10 10]{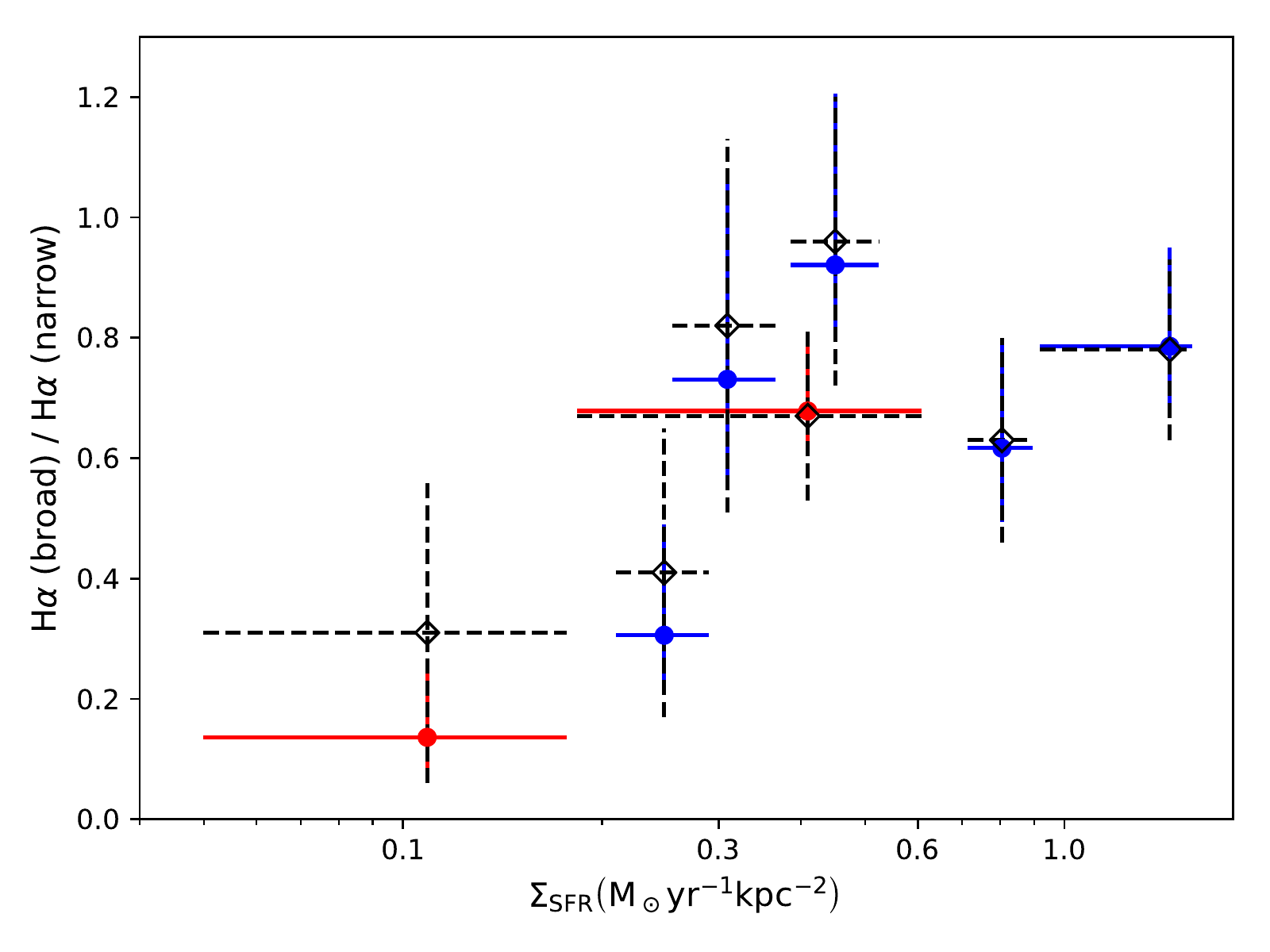}
\caption{Same as Figures \ref{fig:sfrsd_fbroad} and \ref{fig:sfrsd_sig}, but with the estimates from forward modelling over-plotted in black. \label{fig:forward_modelling_estimates}}
\end{figure*}

The results of the forward modelling are summarised in Figures \ref{fig:threshold_recovery_test} and \ref{fig:forward_modelling_estimates}. Figure \ref{fig:threshold_recovery_test} illustrates the mapping between fit space and intrinsic space for two data stacks - one with low \sfrsd\ (top row) and one with high \sfrsd\ (bottom row). In each row, the left hand panel represents fit space and the right hand panel represents intrinsic space. The yellow, orange and red shaded regions in each left hand panel show the 1$\sigma$, 2$\sigma$ and 3$\sigma$ levels of the joint posterior PDF between the fit $\sigma_b$ and the fit BFR for the relevant data stack. The dots in both left hand panels show the fit $\sigma_b$ and BFR values for the complete set of synthetic stacks, but the sizes and colors of the dots differ between the panels to indicate the weighting assigned to each synthetic spectrum in the mapping for the relevant data stack (larger, darker dots indicate higher weighting). The right hand panels illustrate where the highest weighted synthetic stacks lie in the intrinsic \mbox{$\sigma_b$ - BFR} space. 

Figure \ref{fig:forward_modelling_estimates} shows how the estimated intrinsic BFR and $\sigma_b$ values for the data stacks compare to the best fit values. The errors on the estimated intrinsic values are calculated using the weighted standard deviation. For all but the lowest \sfrsd\ stack, the best fit and intrinsic parameters match extremely well, confirming that the outflow parameters are tightly constrained in this regime. For the below median \sfrsd\ stack, the best fit and intrinsic parameters are consistent within the 1$\sigma$ errors, but there is some evidence that the best fit BFR may be under-estimated by a factor of $\sim$2, and the best fit $\sigma_b$ may be under-estimated by $\sim$~30~\kms. Overall, the forward modelling suggests that the $\sigma_b$ may flatten at low \sfrsd, and confirms that the BFR declines at \sfrsd~$<$~0.3~\myrkpc, but indicates that the rate and shape of this decline is uncertain.

\bibliography{../mybib}

\end{document}